\documentclass[letterpaper,twocolumn,10pt]{article}
\usepackage{usenix2019_v3}

\usepackage{tikz}
\usepackage{amsmath}
\usepackage{amssymb}
\usepackage{algorithmic}
\usepackage{graphicx}
\usepackage{textcomp}
\usepackage{xcolor}
\def\BibTeX{{\rm B\kern-.05em{\sc i\kern-.025em b}\kern-.08em
    T\kern-.1667em\lower.7ex\hbox{E}\kern-.125emX}}
\usepackage{multirow,threeparttable,colortbl}

\usepackage{makecell}
\usepackage{wrapfig}

\definecolor{babypink}{rgb}{0.96, 0.76, 0.76}
\definecolor{bananamania}{rgb}{0.98, 0.91, 0.71}
\definecolor{cambridgeblue}{rgb}{0.64, 0.76, 0.68}

\usepackage{pifont}
\newcommand{\cmark}{\textcolor{green!80!black}{\ding{51}}}

\usepackage{appendix}

\usepackage{enumitem}

\usepackage{url}

\usetikzlibrary{arrows.meta}

\usepackage{hyperref}

\usepackage{wrapfig}
\usepackage{float}

\usepackage{mathtools}

\usepackage{listings}
\usepackage{blindtext}
\usepackage[ruled,linesnumbered]{algorithm2e}
\usepackage{etoolbox,xstring,mfirstuc,textcase}
\usepackage{booktabs} 
\usepackage{listings}

\usepackage{comment}
\usepackage{subcaption}

\begin{document}

\date{}

\title{\Large \bf Decentralized Finance (DeFi): A Survey}


\author{
{\rm Erya Jiang$^*$, Bo Qin$^*$, Qin Wang$^\P$, Zhipeng Wang$^\S$, Qianhong Wu$^\dag$, Jian Weng$^\ddag$} \\ 
{\rm Xinyu Li$^*$, Chenyang Wang$^*$, Yuhang Ding$^*$, Yanran Zhang$^*$}\\
\normalsize $^*$Renmin University of China, China\\ 
\normalsize $^\P$University of New South Wales, Australia\\
\normalsize $^\S$Imperial College London, UK\\
\normalsize $^\dag$Beihang University, China\\
\normalsize $^\ddag$Jinan University, China
} 

\maketitle

\begin{abstract}
Decentralized Finance (DeFi) is a new paradigm in the creation, distribution, and utilization of financial services via the integration of blockchain technology. Our research conducts a comprehensive introduction and meticulous classification of various DeFi applications. Beyond that, we thoroughly analyze these risks from both technical and economic perspectives, spanning multiple layers. We point out research gaps and revenues, covering technical advancements, innovative economics, and sociology and ecology optimization.
\end{abstract}

\section{Introduction}
With blockchain's rise, Decentralized Finance (DeFi)~\cite{werner2022sok} has emerged as a disruptive financial paradigm in the middle of 2020 (a period known as the \textit{DeFi summer}), challenging traditional finance~\cite{qin2021cefi}. DeFi utilizes blockchain for creating, distributing, and utilizing financial services~\cite{schueffel2021defi}, surpassing traditional finance in various aspects:

\begin{itemize}[nosep]
    \item \textit{Trustless.} DeFi protocols eliminate centralized intermediaries like brokerages, banks, and insurance companies, which come with defects such as high costs, cumbersome processes, account opening restrictions, e.g., Know Your Customer~(KYC), and lack of transparency.
    \item \textit{Non-human intervention.} DeFi's trading rules are pre-written, making automation and immutability key features~\cite{li2022smart} while running on-chain that reduces counterparty risk and eliminates the single point of failure.
    \item \textit{Maximal availability.} Most DeFi products have no downtime, enabling 24/7 financial services to everyone. 
    \item \textit{Borderless.} DeFi enables global accessibility without being restricted by national boundaries, allowing anyone from anywhere including those from low economic levels or underserved regions to leverage its services.
    \item \textit{Permissionless.}  Deployed in a decentralized manner across peer-to-peer (P2P) networks, opens new opportunities for organizations, e.g., DAOs~\cite{yu2023leveraging} to areas previously accessible only to licensed institutions.
    \item \textit{Extensibility.} DeFi's open-source nature encourages user contributions, facilitating the emergence of novel financial concepts such as flash loans~\cite{wang2020towards, qin2021attacking}.
\end{itemize}

As of Aug. 2023, the total locked value (TLV)\footnote{Data source: \href{https://defillama.com}{https://defillama.com} [Aug. 2023], marked as $\mathsf{\#DefiLlama}$.} in DeFi markets has reached US\$40.257b, following a peak value of US\$253b in Dec. 2021 (also claimed in~\cite{zhou2022sok}). This substantial investment has sparked numerous innovations, including decentralized exchanges (DEXs, e.g., Uniswap~\cite{adams2021uniswap}, dYdX~\cite{juliano2018dydx}), lending (Compound~\cite{compund}, Aave~\cite{aave}), yield aggregators (Convex~\cite{convex}, Harvest~\cite{harvest}), liquid staking (Lido~\cite{lido}, Rocket Pool~\cite{rocket}), and various other developments~\cite{werner2022sok}.

\smallskip
\noindent\textbf{Previous investigations.} We highlight several recent studies that have elegantly reviewed DeFi-related concepts. Wener et al. \cite{werner2022sok} conducted the first systematic studies focusing on general DeFi protocols, covering layer-based protocols and services. Moin et al. \cite{moin2020sok} classified major stablecoin designs, while Zhao et al. \cite{zhao2021understand} specifically explored algorithmic stablecoins. Bartoletti et al. \cite{bartoletti2021sok} introduced lending protocols using a formal model that describes their transitions as a state machine reflecting user interactions. Xu et al. \cite{xu2022short} presented a general business model for a small corpus of DeFi protocols, including protocols for loanable funds, DEXs, and yield aggregators. Li et al. \cite{li2022sok} describe the picture of confidential smart contrast that can be used for DeFi privacy. Xu et al. \cite{xu2023sok} comprehensively reviewed DEXs and their corresponding automated market maker (AMM) protocols. Erinle et al.~\cite{erinle2023sok} provided a comprehensive overview of cryptocurrency wallets that support DeFi services. Lastly, Zhou et al. \cite{zhou2022sok} thoroughly investigated a range of attacks and incidents in the DeFi space. Additionally, a series of research works have drawn their focus on MEV extractions \cite{qin2022quantifying,yang2022sok} and frontrunning attacks \cite{eskandari2020sok,zhou2021just,wang2022exploring}. 

\begin{figure}[ht]
    \centering
    \includegraphics[width=1\linewidth]{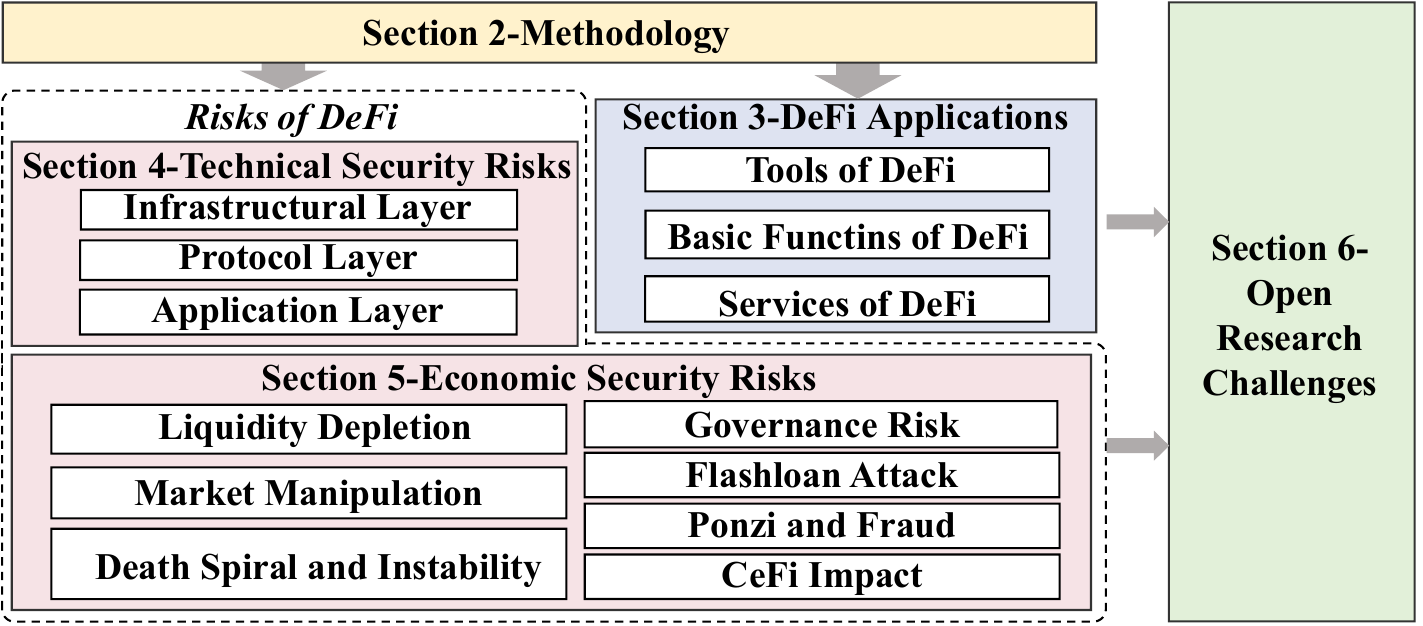}
    \caption{Paper Structure}
\label{fig:structure}
\end{figure}

\textit{{This paper}} follows the burgeoning prosperity of DeFi and extends its research horizons. We explore the mechanisms of DeFi apps and investigate the security risks from technological and economic perspectives (cf. Figure~\ref{fig:structure}). In particular,

\begin{itemize}[nosep]
    \item \textit{We conduct a comprehensive statistical analysis of the literature in the DeFi field.} According to the analytical methods (Section~\ref{sec-method}), we obtain more than 10,000 DeFi-related research articles and conduct qualitative and quantitative analyses.
    \item \textit{We propose a DeFi classification frame based on the complexity of financial services.} The frame  (summarized in Table~\ref{tab-summary}) classifies DeFi applications into three categories (Section~\ref{sec-defiApp}): tool level, basic functionality level, and service level. We detail the structure of DeFi applications and related research insights in each category.
    \item \textit{We discuss the security of DeFi applications from two pillars:}  technical (Section~\ref{sec-technical}) and 
    economic perspectives (Section~\ref{sec-economic}).
    Our discussions are grounded in relevant academic papers and real-world incidents, outlining a broad spectrum of DeFi risks, possible losses, implementations, and possible defenses (summarized in Table~\ref{table:Tech-attack}).
    \item \textit{We provide information on the gap between existing DeFi realizations and the ideal state.} We conclude by proposing technological, sociological, and economic research directions (Section~\ref{sec-direction}). 
\end{itemize}

\section{Methodology}
\label{sec-method}
We provide the analytical approach of statistical literature analysis and outline the corresponding modeling process. The results based on quantitative and qualitative analysis will be presented in the following sections.

Our analysis started with retrieving academic papers and research articles related to DeFi from IEEE Xplore, ACM, and Springer. After keyword screening and manual checking, we obtained a total of 15,598 relevant articles.

\begin{figure}[!hbt]
    \centering
    \includegraphics[width=1\linewidth]{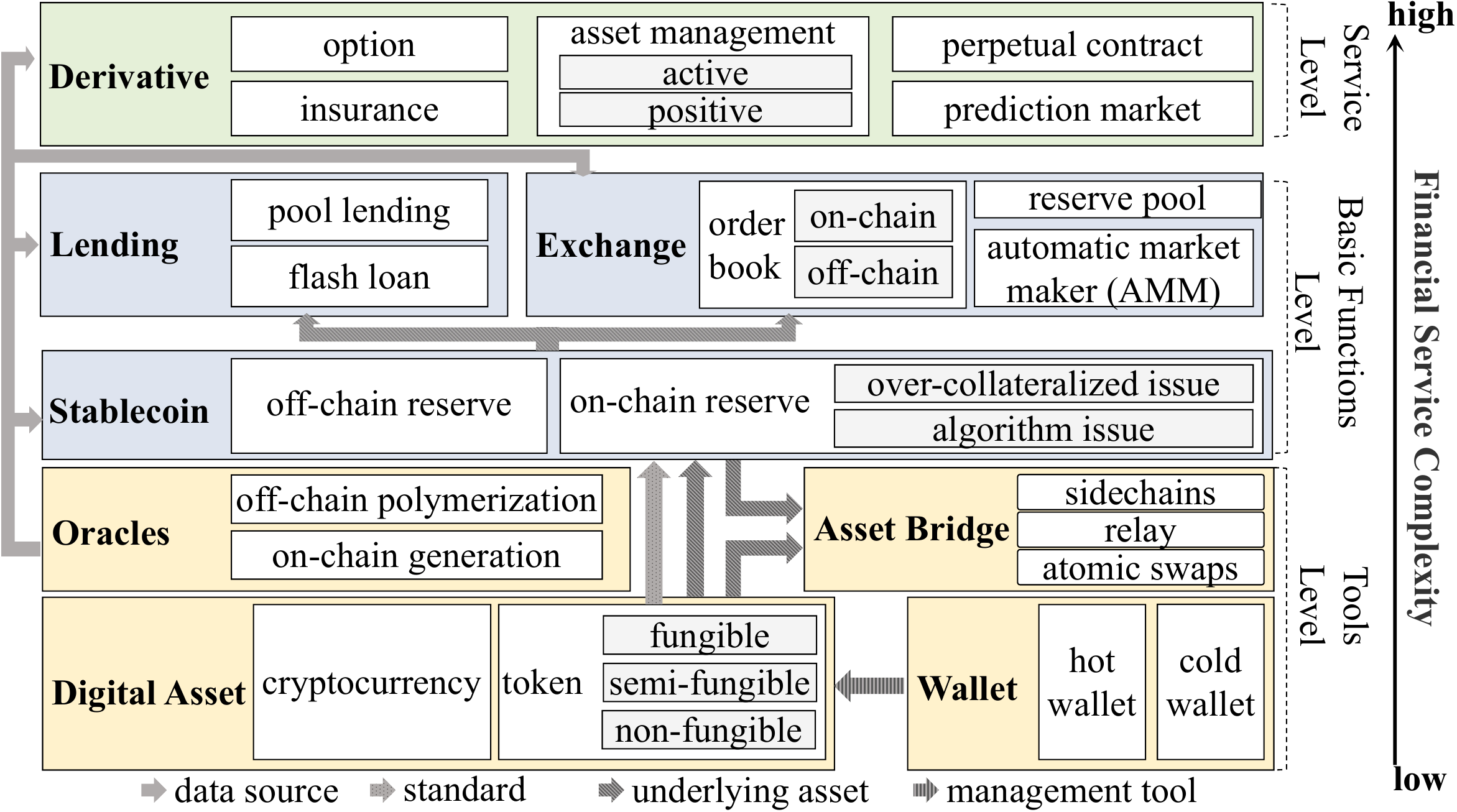}
    \caption{DeFi Ecological Structure}
    \label{fig:deFi-ecological-structure}
\vspace{-0.1in}
\end{figure}

In reviewing the existing literature, we realized that an effective way to classify the themes of existing works is based on the financial complexity of the DeFi application, as shown in Figure~\ref{fig:deFi-ecological-structure}. According to this classification, DeFi applications can be classified as \textit{digital assets}, \textit{wallets}, \textit{oracles} and asset \textit{bridges} at infrastructural level, \textit{stablecoins}, \textit{lending}, and \textit{exchanges} at functional level, and diverse \textit{derivatives} at service level. In addition, we have screened and classified the literature that offers DeFi security risk solutions.

We utilized quantitative techniques including employing descriptive statistics and other tools to establish relationships, trends, and statistical significance to analyze the data. We also employed qualitative analysis to complement the quantitative analysis. This allowed us to gain a holistic understanding of the subject matter and capture nuanced insights.

\section{DeFi Applications}
\label{sec-defiApp}
We conducted a statistical analysis on over 10,000 DeFi-apps-related literature pieces that we obtained. 

\smallskip
\noindent\textbf{Observation.} DeFi research literature availability correlates with proximity to blockchain logic. We observe that research on DeFi tools and functionalities closely tied to blockchain emerged earlier and has more quantity. For example, digital assets research, due to the native token-blockchain consensus relationship, is the most abundant in DeFi. 

\noindent\colorbox{cambridgeblue}{\textbf{Insight.}} Initially, DeFi research focused on fundamental tools and logic. This is because establishing a reliable and secure blockchain infrastructure was crucial during the early stages of blockchain and DeFi development.  This involved core technologies like decentralized development, consensus algorithms, network security, and smart contract platforms for automated financial functions.  These studies deepened the understanding of blockchain's underlying logic, resulting in extensive literature.  As the infrastructure was established, exploration and development of complex DeFi tools and applications followed.  Early research mirrored the developmental process of the financial system, which begins with establishing basic tools and infrastructure before expanding to a wider range of products and services.  Similarly, DeFi progressed from infrastructure construction to the research and development of diverse tools and applications.

\smallskip
\noindent\textbf{Observation.} DeFi research has seen a significant increase, especially in 2022 and 2023. These papers explore previously unexplored aspects of DeFi, such as DeFi derivatives. 

\noindent\colorbox{cambridgeblue}{\textbf{Insight.}} The research boom in DeFi since 2022 is attributed to several factors, including the maturation and expansion of the DeFi market, technological innovations, and increased involvement of institutions and enterprises. As the DeFi market matures, there is a rising demand for diverse and complex DeFi products, including derivatives. The validation of DeFi's potential through industry practices has also played a role. Introduction of new technologies like Layer-2 scaling solutions, cross-chain technologies, and privacy protection has opened up new possibilities. Additionally, the participation of enterprises and the regulatory measures and resources they bring have further fueled research in DeFi.

\begin{figure}[!hbt]
    \centering
    \includegraphics[width=0.95\linewidth]{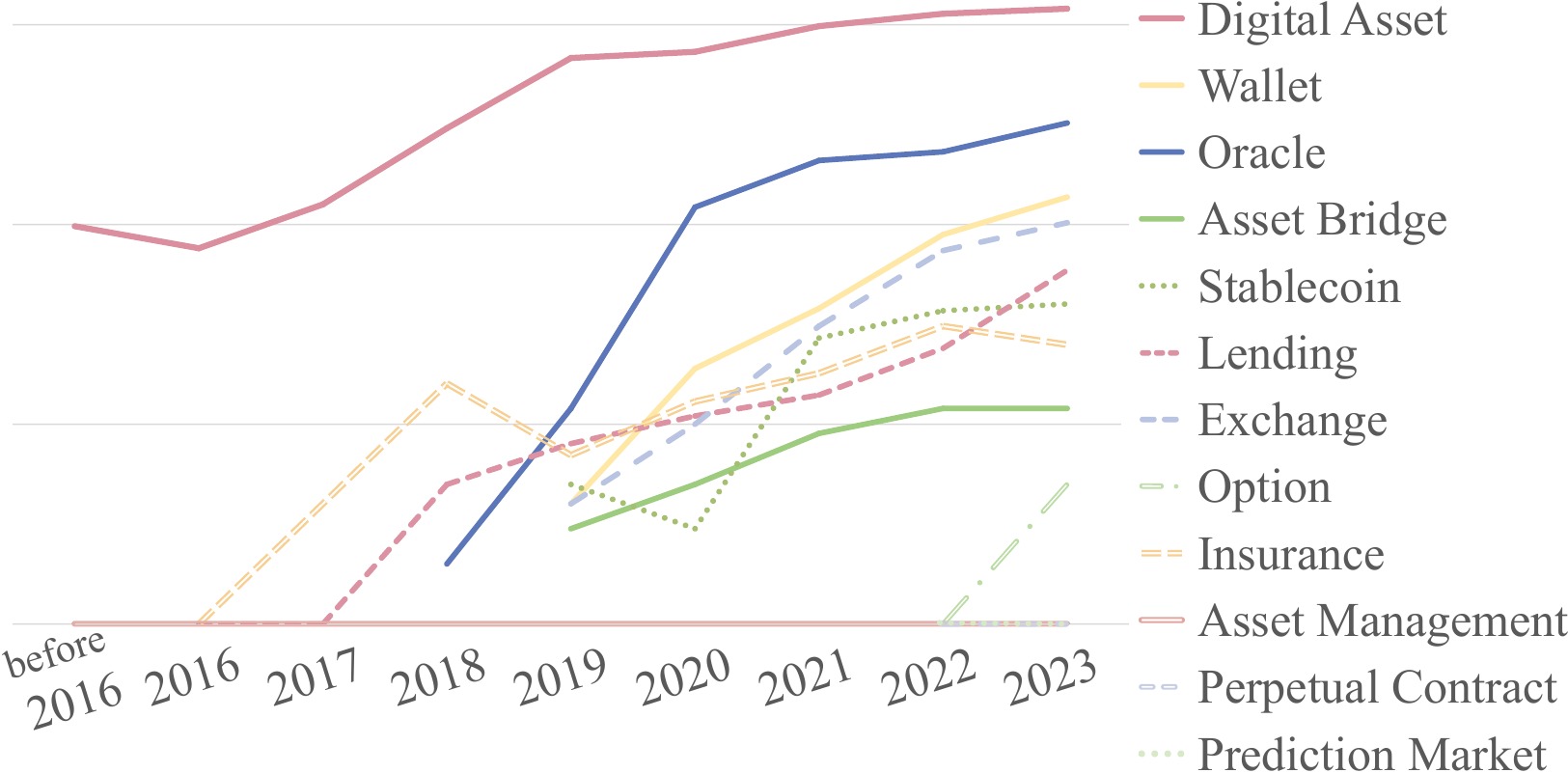}
    \caption{Trend of Research Literature on DeFi Applications (Log10 Scale): We illustrate the trends of literature related to different DeFi apps over time. Notably, literature on asset management, prediction markets, and perpetual contracts is limited, leading to overlapping lines with the horizontal axis.}
    \label{fig:year-researched}
\end{figure}

\smallskip
\noindent\textbf{Observation.} Technical solutions constitute the main type of literature in the DeFi field. 

\noindent\colorbox{cambridgeblue}{\noindent\textbf{Insight.}} Technical solutions being the focus of researchers and practitioners is inherent as DeFi is driven by technology practice and application. They play a key role in transforming theories into practical tools and platforms.

\begin{figure}[!hbt]
    \centering
    \includegraphics[width=\linewidth]{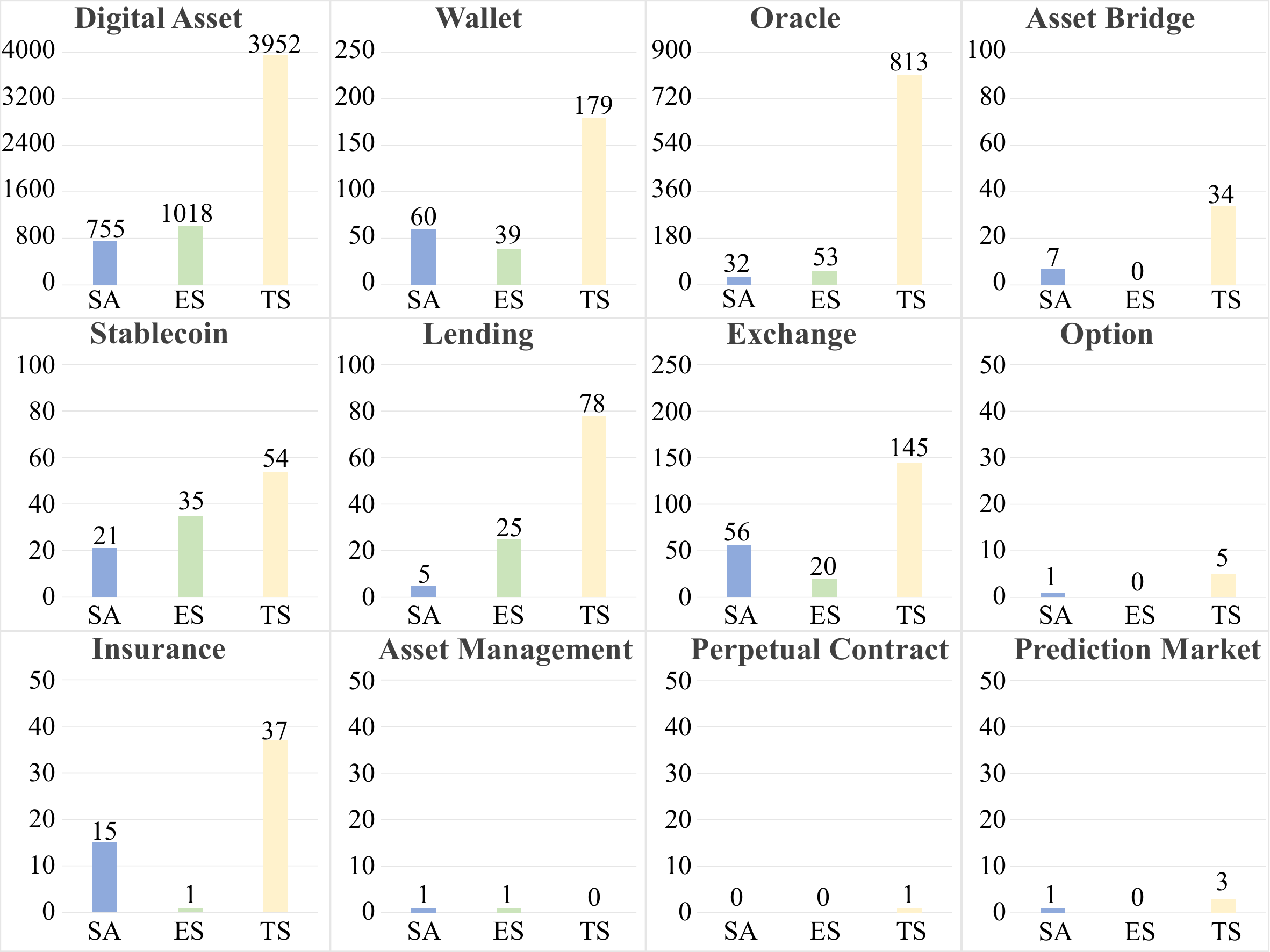}
    \caption{Publications of Literature Type: We classify the literature into three categories: review articles \textit{(blue)}, empirical studies \textit{(green)}, and technical solutions \textit{(yellow)}. We summarize the distribution of different types of literature.}
    \label{fig:literature-type}
\end{figure}

\smallskip
\noindent\textbf{Observation.} Review articles often assess from a single disciplinary perspective, restricting comprehensive evaluations. 

\noindent\colorbox{cambridgeblue}{\textbf{Insight.}} DeFi encompasses multiple disciplines, including economics, information security, law, and more. When review articles lean towards a single disciplinary perspective, they tend to be influenced by specific viewpoints and limitations. This ultimately stems from the fact that authors may be influenced by their own disciplinary viewpoints, leading to a lack of understanding of other disciplines. This reflects the insufficient depth of collaboration among different disciplinary talents, highlighting the need to cultivate interdisciplinary talents with a comprehensive perspective.

\smallskip
\noindent\textbf{Observation.} Empirical research lacks a macro perspective and broader significance in investigating DeFi apps. Apart from empirical research in the field of digital assets, which includes market data analysis, other areas rely more on case studies investigating specific projects, resulting in a lack of macro perspectives and broader empirical investigations.

\noindent\colorbox{cambridgeblue}{\textbf{Insight.}} Due to the novelty of DeFi, empirical research in this domain faces challenges.  Inconsistent definitions and unclear models complicate empirical investigations.  Privacy and data protection concerns surrounding DeFi apps make it difficult to acquire sufficient data for macro perspectives and broad empirical studies.  The rapid emergence of new DeFi apps and models also poses challenges in keeping up with the pace of innovation and conducting comprehensive research.  Consequently, empirical research in DeFi tends to focus more on specific case studies rather than broader investigations.


\subsection{Tools of DeFi}
\label{sec-infraApp}

\noindent\textbf{Digital Asset.} Native cryptocurrencies and derivative tokens constitute the money flowing in DeFi.

\noindent\textit{\textbf{Native cryptocurrency.}} Native cryptocurrency refers to the primary digital asset of a blockchain. Prominent examples include Bitcoin~\cite{nakamoto2008bitcoin}, Ethereum \cite{buterin2014next}, Litecoin~\cite{Litecoin}, Monero~\cite{alonso2020zero,sun2017ringct,hinteregger2018empirical}, and Zcash~\cite{hopwood2016zcash,kappos2018empirical}, all operating on standalone blockchains and incentivized within their respective economies. These cryptocurrencies can be directly transferred on the chain that hosts them and used for paying transaction fees, serving various functions within its ecosystem.

\noindent\textit{\textbf{Derivative Token.}} Many DeFi apps issue tokens representing the ownership of the assets corresponding to the app. Additionally, DeFi app-issued tokens can perform in lending~\cite{aave}, staking~\cite{rocket}, and insuring~\cite{nexus2020}, and be empowered to participate in governance. Tokens in physical form represent ownership of real-world assets like real estate and collectibles. 

Ethereum has a wide range of token standards for various DeFi apps. The widely known ERC-20 standard is fungible and interchangeable and used for currencies, voting tokens, and pledge tokens. Other Ethereum token standards are tailored for specific scenarios, such as non-fungible token standard ERC-721 used in artwork and digital collections, and semi-fungible token standard ERC-1155 utilized in GameFi and copyright. ERC-998 is designed to combine ERC-20 and ERC-721, enabling compatibility and interoperability. Other blockchain platforms like Binance Smart Chain (BSC), Avalanche, and Bitcoin have also introduced their own token standards, such as BEP-20, BEP-721, ARC-721, and BRC-20, following similar rules to the ERC standards.

There is controversy regarding the classification of DeFi tokens as currencies, commodities, or securities. Some DeFi tokens exhibit currency attributes, possessing wide usability and circulation, while others may be classified as securities due to their characteristics of representing ownership, dividends, or investment returns. The classification also relies on regulatory standards, which are varied across regions and continuously adjusted given the technical complexity of DeFi.

The digital asset research field within DeFi emerged early and has a wealth of literature. Quantitative and predictive research, specifically focused on the cryptocurrency market, is a significant area of study. Quantitative research employs historical market data to develop trading strategies and algorithms based on technical analysis indicators, statistical models, and machine learning methods. Predictive research, on the other hand, involves constructing forecasting models using time series analysis, machine learning, and deep learning techniques. Notably, the integration of machine learning and deep learning, along with comprehensive consideration of market characteristics and risks, are prominent features within this research field.


\smallskip
\noindent\textbf{Wallet.} A wallet is a tool for managing the keys and addresses of digital asset holders. Wallets serve to interact with the blockchain instead of storing on-chain assets.  In DeFi, users can manage multiple accounts from a single wallet.

Typically, a wallet has three basic functions: recording, receiving, and transferring currencies. With the development, its functions have evolved from simple transfers to encompass multi-chain management, asset custody, and other scenarios. Numerous academic research and industry examples on the functional expansion of wallets exist. Software wallets like imToken \cite{imToken}, Bip \cite{Bip}, Wetez \cite{Wetez}, TrustWallet \cite{TrustWallet}, and hardware wallets like Ledger \cite{Ledger} and Trezor \cite{Trezor} have explored and implemented these functionalities. In the industry, multi-chain wallets are typically developed by creating interfaces for different blockchains. Some multi-chain wallets have even introduced ``flash exchange'' functionality, utilizing exchange rates as a medium for transactions.



Wallet security is a critical consideration, encompassing key preservation, recovery procedures, and risk mitigation. Cold wallets offer physical isolation but carry the risk of loss, which has led to the development of hot and non-custodial wallets. Multi-factor authentication, such as biometrics and behavioral features, has been implemented to enhance security. Secret sharing and Trusted Third Party (TTP) verification have also been employed to strengthen security of key recovery. 

\noindent\textbf{Oracle.}
The execution of smart contracts requires meeting conditions specified in the contracts, while also requiring support from external data. Oracle provides external data sources for smart contracts on the blockchain, supplying them with data information. As of Oct. 2023, the total value of all oracles reached US\$25b, with Chainlink holding nearly half of the market among more than 40 different oracles ($\mathsf{\#DefiLlama}$). 

To ensure trustworthy on-chain data, the accuracy of data is the main concern~\cite{mackinga2022twap}. We have observed that review articles in this field generally classify oracles based on their operational mechanisms, which can be broadly categorized into provider identity-based and voting-based oracles. Identity-based oracles involve specific implementation methods such as setting whitelists/blacklists, incorporating identity verification in transport layer protocols, or using machine learning techniques to identify reliable and cost-effective oracles~\cite{taghavi2023reinforcement}. MakerDao~\cite{makerdao}  is an example of this type of oracle. On the other hand, voting-based oracles incentivize providers to exhibit economically rational behavior and provide accurate data through monetary rewards and penalties. Implementation methods for voting-based oracles include peer-to-peer prediction~\cite{cai2022truthful,goel2019decentralized}, reputation mechanisms~\cite{breidenbach2021chainlink}, game-theoretic approach for price data verification~\cite{nest2023,nest}, and others.


While different approaches have been implemented, challenges remain in ensuring data security. Identity-based oracles are vulnerable to single points of failure and bribery attacks. Voting-based oracles have limited applicability due to the need for data verification, which restricts them to publicly accessible information. These oracles also face challenges like data latency and high verification costs. Furthermore, the timeliness and freshness of time-sensitive data are often overlooked in current research, creating a dilemma in balancing security and timeliness.

\noindent\textbf{Asset Bridge. }
Heterogeneous blockchains present a challenge to achieving smooth interoperability in DeFi. Asset bridge is a solution for asset transfer and interoperability between different blockchains. In the days that have passed in 2023, the average daily trading volume of the asset bridge exceeded \$214 million, with the highest daily trading volume reaching \$1.232b ($\mathsf{\#DefiLlama}$), reflecting active trading activity in the business. The functioning of an asset bridge varies depending on the implementation. Atomic swaps allow the direct exchange of cryptocurrency across blockchains~\cite{wang2023exploring}. Ripple introduced the InterLedger protocol (ILP) in 2012, facilitating cross-ledger interactions through third-party notaries. Pegged sidechains were proposed by the Bitcoin Core development team in 2014. 
Interoperability platforms such as Cosmos~\cite{kwon2019cosmos}and Polkadot~\cite{wood2016polkadot} realize cross-chain communication and interaction through relay chains or side chains. In 2015, Joseph Poon and Thaddeus Dryja conceptualized the Bitcoin Lightning Network. In 2016, BTC-Relay~\cite{btcrelay}, a cross-chain solution based on a relay chain, achieved one-way cross-chain connectivity between Ethereum and Bitcoin~\cite{relay2016frequently}. Vitalik Buterin~\cite{buterin2016chain}'s effort provided an in-depth analysis of blockchain interoperability issues. Notable cross-chain DeFi applications include Thorswap~\cite{thor} and Chainswap~\cite{9996367}.

\subsection{Basic Functions of DeFi}
\label{sec-funcApp}

\noindent\textbf{Stablecoin.} The prices of cryptocurrencies are highly volatile, but stablecoins offer price stability as they are pegged to fiat currencies, which is exactly why stablecoins were born. As a foundational currency, stablecoins support liquidity pools, lending, insurance, and other financial activities~\cite{catalini2022some}, mitigating the risks associated with market fluctuations. As of Oct. 2023, the stablecoin market contains a total market capitalization of over \$120b and over 100 projects ($\mathsf{\#DefiLlama}$).


Stablecoins circulation involves reserve, insurance, and other essential links.
Methods of forming stablecoins include off-chain reserves, such as USDT~\cite{tether}, USDC~\cite{usdc}, and GUSD~\cite{gusd}, on-chain collateralization like Dai~\cite{makerdao} and LUSD~\cite{liquity}, and algorithmic stablecoins without collateralization such as AMPL~\cite{kuo2019ampleforth}, Basis~\cite{al2017basis}, FRAX~\cite{kazemian2022frax}, and UST~\cite{kereiakes2019terra}. Among them, we have found that algorithmic stablecoins are a controversial object of research. While having the advantages of high transparency and low/no collateral rates, multiple algorithmic stablecoin crashes, including Luna-UST collapse in 2022~\cite{briola2023anatomy}, have cast a shadow over this solution. To address this challenge, Klages Mundt et al.~\cite{klages2021stability} propose modeling-based approaches to enhance stablecoin design and resilience, ensuring price stability even amidst market shocks. Fu et al.~\cite{fu2023rational} propose a rational Ponzi model to analyze the sustainability of algorithmic stablecoins.


A critical issue we identified in stablecoin research is the lack of widely accepted definitions for stablecoins. Existing literature often lacks clarity in defining what precisely constitutes a stablecoin. While some review articles discuss various implementation approaches, they do not provide a definitive core definition. Many papers focus on highlighting the desired characteristics and advantages of an ideal stablecoin, without delving into a concrete definition. Only a few studies examine stablecoin definitions from a legislative perspective or attempt to model specific types of stablecoins.

\smallskip
\noindent\textbf{Lending.} DeFi lending abandons the centralized credit assessment framework but relies on recognized collateral for pooling liquidity, enabling low-cost lending and arbitrage, and improving the transferability of debt holdings. DeFi lending has a large market, with a TLV of \$14.782b and 300+ Apps ($\mathsf{\#DefiLlama}$). It allows borrowers to engage in trading activities, while lenders can earn additional revenue via collateral rates~\cite{black2019atomic}. The primary motivation for users to use DeFi lending is to obtain participation rewards, such as governance tokens. In extreme cases, investors can form a \textit{borrowing spiral}~\cite{pham2022analysis} or \textit{leverage spiral}~\cite{saengchote2022decentralized} to maximize benefit.


DeFi lending apps typically involve collateralization, lending, and liquidation. 
Based on such a model, apps like Compound~\cite{compund} and AAVE~\cite{aave}, enable over-collateralized, trust-less DeFi lending. But out of the demand for low/zero collateralization and regulatory requirements, undercollateralized lending is born, building credit on the blockchain and setting the constraints for using the borrowed assets~\cite{wang2022speculative}. Solutions based on credit assessment models~\cite{uriawan2021collateral,xie2022towards} and incentive punishment mechanisms~\cite{hassija2020secure} are proposed, respectively. In the industry, TrueFi~\cite{truefi},  DeFi Passport~\cite{DeFiPassport}, and CreDA~\cite{creda2022} have carried out the practice of on-chain credit assessment.


Flash loans are DeFi's innovative non-collateralized lending tool and have various use cases, such as arbitrage, collateral swapping, and self-liquidation~\cite{wang2020towards}. Flash swaps provide similar services to flash loans within DEXs. Both flash loans and flash swaps leverage the atomicity of transactions, utilizing optimistic transfers that enable collateral-free loans or token exchange transactions as long as the loan is repaid by the end of the block (illustrated in Figure~\ref{fig:fls}).

\smallskip
\noindent\textbf{Exchange.} In traditional exchanges, market makers summarize trades based on the seller's request and the buyer's offer on the order book. Decentralized exchanges (DEXs) decentralize aggregation, clearing, and market making through blockchain~\cite{barbon2021quality, berg2022empirical}. More than 1000 apps have made DEXs the most abundant application type in DeFi, with a TLV of US\$11.498b ($\mathsf{\#DefiLlama}$).

DEX can be divided into different models based on the implementation of trading pair discovery and order matching~\cite{heimbach2021behavior}. 
These models include on-chain order book model as implemented by Stellar, off-chain order book model as implemented by 0x~\cite{warren20170x}, AirSwap~\cite{oved2017swap}, IDEX~\cite{IDEXwhite} and dYdX~\cite{juliano2018dydx,starkex}, and non-order book model which is one of the most important innovations of DeFi. Methods of no-order book model, including reserve pool (implemented by KyberNetwork~\cite{luu2017kybernetwork}) and algorithms of AMM like constant mean (adopted by Balancer~\cite{martinelli2019non}), constant product (adopted by Uniswap~\cite{adams2021uniswap}), dynamic weighting(adopted by Bancor~\cite{hertzog2017bancor}) and mixed-function algorithm(adopted by the Curve Finance~\cite{egorov2021automatic}), have been implemented in the industry. This is also one of the hottest research areas, including reviews that categorically evaluate different implementations~\cite{xu2023sok,mohan2022automated,angeris2020improved} and specific algorithms that have been put into practice in the industry. We observe that empirical research is a kind of literature that is relatively lacking. There are some case studies on individual algorithms like \cite{egorov2019stableswap}, but there is a lack of empirical studies based on extensive data.

\begin{table*}[!bth]
\caption{DeFi Construction and Classification}\label{tab-summary}
\vspace{-0.9em}
\begin{center}
\begin{threeparttable}

\resizebox{0.99\textwidth}{!}{
\begin{tabular}{lr c|cccccc|ccc}
\toprule
\multicolumn{1}{c}{} 
& \multicolumn{1}{c}{} 
& \multicolumn{1}{c}{} 
& \multicolumn{6}{c}{\textbf{Feature}} 
& \multicolumn{3}{c}{\textbf{Property}} 
\\
\cmidrule(lr){4-9}
\cmidrule(lr){10-12}

& 
\rotatebox{0}{\textbf{Project}}& 
\rotatebox{0}{\textbf{Type}} & 
\rotatebox{60}{\textit{Trust Model}}  &
\rotatebox{60}{\textit{Connected Chains}} &
\rotatebox{60}{\textit{Centralization }} & 
\rotatebox{60}{\textit{Anonymous}} &
\rotatebox{60}{\textit{Tokenization}} &
\rotatebox{60}{\textit{Technique}} &
\rotatebox{90}{\textit{Stability}} & 
\rotatebox{90}{\textit{Scalability}}  & 
\rotatebox{90}{\textit{Complexity}} 
\\   
\midrule 
\multirow{7}{*}{\rotatebox{90}{\textbf{Digital Assets}}} & \cellcolor{cambridgeblue} Bitcoin\cite{nakamoto2008bitcoin}  &  \multirow{5}{*}{Native Crypocurrency}  & - & One &  - & M. & -  & BC & -  & - & - \\

& \cellcolor{cambridgeblue} Ethereum\cite{buterin2014next}   & & - & One  &  - & M. & -  & BC & -  & - & - \\

& \cellcolor{cambridgeblue} Litecoin  & & - & One  &  - & M. & -  & BC & -  & - & - \\

& \cellcolor{cambridgeblue} Monero\cite{Monero}  & & - &  One & - & H.   & - & Ring-sig, Stealth Address & -  & - & - \\

& \cellcolor{cambridgeblue} Zcash\cite{Zcash}  & & - &  One  & - & H. & -  & zk-Snark, Multi-sig & -  & - & - \\

\cmidrule{2-3}

& \cellcolor{cambridgeblue} ERC-20\cite{erc20eips}  & \multirow{1}{*}{Token Standard} & - & One &  Up to Apps & M. & - & SC & -  & - & - \\

 
 

\midrule

\multirow{6}{*}{\rotatebox{90}{\textbf{Wallets}}} & \cellcolor{cambridgeblue}  Ledger\cite{Ledger}  &  \multirow{2}{*}{Cold Wallet} & - & Multiple &  - & M. & -  & TEE & -  & - & - \\

& \cellcolor{cambridgeblue} Trezor\cite{Trezor}  & & - & Multiple &  - & H. & -  & Multi-sig, 2FA & -  & - & - \\

\cmidrule{2-3}
& \cellcolor{cambridgeblue} Metamask\cite{metamask}  & \multirow{3}{*}{Hot Wallet}& - & Multiple & - & H. & -  & Offline Storage & -  & - & - \\


& \cellcolor{cambridgeblue} Zengo\cite{zengo} & & TTP & Multiple & - & M. & -  & MPC-TSS, 3FA & -  & - & - \\

& \cellcolor{cambridgeblue} Argent\cite{Argent} & & TTP & Multiple & - & H. & -  & Multi-sig, 2FA & -  & - & - \\


\midrule

\multirow{4}{*}{\rotatebox{90}{\textbf{Oracle}}} & \cellcolor{cambridgeblue}  MakerDao\cite{makerdao}  &  \multirow{1}{*}{Alliance Oracle, Stablecoin}& TTP & One &  DAO & M. & Dai, MKR  & Allowlist & H.  & M. & L. \\

\cmidrule{2-3}

& \cellcolor{cambridgeblue} Chainlink\cite{breidenbach2021chainlink} & \multirow{1}{*}{Off-chain Input}& - & Multiple & - & M. & LINK & Reputation, Staking & H. & H. & M. \\

\cmidrule{2-3}
& \cellcolor{cambridgeblue}  NEST\cite{nest}  & \multirow{1}{*}{Fact Generation} & - & One &  - & M. & QP Token  & Game Theory & H.  & M. & H. \\

\midrule

\multirow{6}{*}{\rotatebox{90}{\textbf{Bridges}}} & \cellcolor{cambridgeblue}  Cosmos\cite{kwon2019cosmos}  &  \multirow{3}{*}{BC Engine} & TTP & Multiple & Hub & H. & ATOM  & IBC & -  & H. & L. \\

& \cellcolor{cambridgeblue}  Polkadot\cite{wood2016polkadot}  & & TTP & Multiple & Validator & M. & DOT  & Parachain & -  & H. & L. \\


\cmidrule{2-3}
& \cellcolor{cambridgeblue}  BTC-relay\cite{btcrelay} & \multirow{1}{*}{Relay}& TTP & Two &  Mainchain & M. & ETH-BTC  & SPV & -  & M. & M. \\


\cmidrule{2-3}

& \cellcolor{cambridgeblue} Polygon Bridge\cite{polygon}  & \multirow{2}{*}{DApps Based}& - & Multiple  &  - & M. & POL  & Lock\&Mint & -  & - & - \\
 
& \cellcolor{cambridgeblue} ThorSwap\cite{thor}  & & - & Multiple(Cosmos)  &  - & M. & RUNE  & Third Party Chains, LP & -  & - & - \\

\midrule

\multirow{7}{*}{\rotatebox{90}{\textbf{Stablecoins}}} & \cellcolor{cambridgeblue}  Tether\cite{tether}  &  \multirow{1}{*}{Off-chain Reserve} & - & Multiple &  Issuer & L. & USDT & - & H.  & H. & L. \\


\cmidrule{2-3}

& \cellcolor{cambridgeblue} Liquity\cite{liquity}  & \multirow{1}{*}{Over-collateral} & - & One & - & M. & LQTY, LUSD & - & M. & M. & M.\\

\cmidrule{2-3}
& \cellcolor{cambridgeblue} Ampleforth\cite{kuo2019ampleforth}  & \multirow{4}{*}{Algorithmic} & - & Multiple &  - & M. & AMPL  & QTM-based Algorithmic & L.  & L. & H. \\

&\cellcolor{cambridgeblue}  Frax Finance\cite{kazemian2022frax}  & & - & One &  - & M. & FRAX, FXS  &  Algorithmic Seigniorage& L.  & L. & H. \\

&\cellcolor{cambridgeblue}  Basis\cite{al2017basis}  & & - & One &  - & M. & BAC, BAS, BAB  & Algorithmic Seigniorage & L.  & L. & H. \\

& \cellcolor{cambridgeblue} Terra\cite{kereiakes2019terra}  & & - & One &  - & M. & UST, LUNA  & Parachain & L.  & M. & H. \\

\midrule

\multirow{5}{*}{\rotatebox{90}{\textbf{Lendings}}} &  \cellcolor{cambridgeblue} AAVE\cite{aave}  &  \multirow{2}{*}{Over-collateral}& - & Multiple &  No & M. & Aave  & - & -  & - & - \\

&  \cellcolor{cambridgeblue} Compound\cite{compund}  & & - & Multiple &  No & M. & COMP  & - & -  & - & - \\

\cmidrule{2-3}
& \cellcolor{cambridgeblue} TrueFi\cite{truefi}  & \multirow{2}{*}{Under-collateral} & TTP & One &  Staker & M. & TRU  & - & -  & - & - \\

&\cellcolor{cambridgeblue}  Maple Finance\cite{maple}  & & TTP & One &  Pool Delegates & L. & MPL & - & -  & - & - \\

\midrule

\multirow{9}{*}{\rotatebox{90}{\textbf{Exchanges}}} & \cellcolor{cambridgeblue}  Stellar\cite{stellar}  &  \multirow{1}{*}{On-chain Orderbook} & - & Multiple &  - & M. & XLM  & Manual Matching & -  & L. & - \\

\cmidrule{2-3}
& \cellcolor{cambridgeblue}  0x\cite{warren20170x} &  \multirow{2}{*}{Off-chain Orderbook} & - & Multiple &  Orderbook & M. & ZRX  & Manual Matching & -  & M. & - \\

& \cellcolor{cambridgeblue}  dYdX\cite{juliano2018dydx}  & & - & Multiple(Cosmos) & Orderbook & M. & DYDX  & Manual Matching & -  & M. & - \\


\cmidrule{2-3}
& \cellcolor{cambridgeblue}  KyberNetwork\cite{luu2017kybernetwork} &  \multirow{5}{*}{Non-orderbook}& - & One &  - & M. &  KNC & Reserve Pool & -  & H. & - \\


&  \cellcolor{cambridgeblue} Bancor\cite{hertzog2017bancor}  &  & - & One  & - & M.  &  BNT & Constant product &  - & H. & -    \\

&  \cellcolor{cambridgeblue} Uniswap\cite{adams2021uniswap} & & - & Multiple  & - & M.  &  UNI & Constant mean &  - & H. & - \\

&  \cellcolor{cambridgeblue} Balancer\cite{martinelli2019non}  & & - & One  & - & M.  &  BAL & Dynamic weight & -  & H. & - \\

&  \cellcolor{cambridgeblue} Curve Finance\cite{egorov2021automatic} & & - & One  & - & M.  &CRV & Hybrid function& -  & H. & - \\

\midrule

\multirow{13}{*}{\rotatebox{90}{\textbf{Derivatives}}} &  \cellcolor{cambridgeblue} Opium\cite{opium2020white}  & \multirow{3}{*}{Option} & TTP & One  & Orderbook & M.  &  OPIUM & Off-chain Orderbook &  - & - & -    \\

& \cellcolor{cambridgeblue}  Opyn\cite{koticha2019opyn}  & & - & One  & - & M.  &
Squeeth  & AMM & -  & - & - \\

& \cellcolor{cambridgeblue}  Hegic\cite{wintermute2020hegic}  & & - & One  & - & M.  &  HEGIC & Peer-to-Pool & -  & - & - \\

\cmidrule{2-3}

 &  \cellcolor{cambridgeblue} Enzyme\cite{Enzyme2023}  &   \multirow{2}{*}{Asset Management} & TTP & One  & DAO/Manager & M.  &  MLN & Active Asset Management & -  & - & -    \\


& \cellcolor{cambridgeblue}  Set\cite{feng2019set} &   & - & One & - & M.  &  - & Passive Strategy, Social Trading &  - & - & -    \\


\cmidrule{2-3}

 & \cellcolor{cambridgeblue} Nexus Manual\cite{nexus2020}  & \multirow{5}{*}{Insurance}  & - & One  & - & M.  &  - & Risk Sharing Pool & -  & H. & -    \\

& \cellcolor{cambridgeblue}   VouchForMe\cite{vouch2018} & & TTP & One  & - & M.  &  - & Social Network Proof & -  & H. & - \\

&  \cellcolor{cambridgeblue}  Augur\cite{peterson2015augur}  & & - & One  & - & M.  &  REP & Prediction Market & -  & H. & - \\

&  \cellcolor{cambridgeblue} CDx\cite{cdx}  & & - & One  & - & M.  &  CDX, Cred & Tokenized CDS & -  & L. & - \\

&  \cellcolor{cambridgeblue} Tokens\cite{koticha2019opyn}  & & - & One  &- & M.  & oToken & Put Option &  - & L. & - \\

\cmidrule{2-3}

& \cellcolor{cambridgeblue}  Augur\cite{peterson2015augur}   & \multirow{2}{*}{Prediction Market}  & - & One  & Staker & M.  &  REP & Voting & -  & - & -    \\

& \cellcolor{cambridgeblue}  Omen\cite{Omen} &   & - & One  & - & M.  &  OWL  & Conditional Tokens & -  & - & - \\
\bottomrule 
\end{tabular} }

\begin{tablenotes}
      \footnotesize
      \item[] $-$ = Does not provide property; 
      \item[] \textbf{Abbr.}: BC = Blockchain; Tx = Transaction;   SC = Smart Contracts; QTM = Quantity Theory of Money;
      \item[] LP = Liquity Pool; IBC = Inter-Blockchain Protocol; CDS = Credit Default Swap; H./M./L. = High/Medium/Low.
     \end{tablenotes}
   \end{threeparttable}
   \vspace{-0.2in}
\end{center}

\end{table*}

\subsection{Services of DeFi}
\label{sec-servApp}

Inspired by traditional derivatives, DeFi offers on-chain options, asset management, and decentralized insurance by replacing traditional processes with on-chain automatic executions~\cite{schueffel2021defi}. New financial derivatives, such as perpetuity contracts and prediction markets, have also emerged.

DeFi derivatives are a recent development in both industry and academia. However, compared to the industry's quick implementation and over US\$1.8b TLV ($\mathsf{\#DefiLlama}$), research on DeFi derivatives is limited, with only a few available papers. Empirical research in this field is almost non-existent, and most review articles focus on discussing the feasibility of DeFi derivatives. They emphasize the benefits compared to traditional solutions but lack evaluations of implemented solutions. This may be related to the novelty of the field.

Additionally, there is a more substantial and earlier body of research related to DeFi insurance, possibly because of the high level of decentralization and collateral involved in the DeFi space, making risk management a crucial concern. Furthermore, frequent DeFi incidents have raised significant concerns about the security of funds. Insurance is seen as a significant tool to mitigate risk and enhance capital security, which is why it received earlier attention and exploration. In contrast, research on other types of DeFi derivatives, such as options, perpetual contracts, and asset management, is relatively scarce. This may be because insurance is a traditional financial instrument with well-established concepts and applications, while other derivatives such as perpetual contracts are still in relatively early stages of development, involving more technical and compliance challenges.

\smallskip
\noindent\textbf{Option.} 
DeFi options enable the buying or selling of an asset at a predetermined price in the future through decentralized platforms. The process is automated by smart contracts and involves two main participants: the buyer and the seller. DeFi markets offer higher efficiency and liquidity compared to traditional options trading. These decentralized options protocols cater to investors seeking high-risk, high-leverage cryptocurrencies for speculation, as well as traders looking for hedging and protection against volatile cryptocurrencies.

The workflow of DeFi options trading is similar to traditional options, with two main players facilitated by smart contracts. Various solutions exist based on the matching process, including off-chain order matching models like Opium~\cite{opium2020white}, where orders are handled off-chain and settled on-chain. AMM mechanisms are implemented by Opyn~\cite{koticha2019opyn}, while Hegic adopts liquid sharing pools~\cite{wintermute2020hegic}. DeFi options encompass standardized European options, some non-standard options, and over-the-counter (OTC) options. Deribit~\cite{deribit}, OKEx~\cite{OKX2022}, and other exchanges have launched standardized options trading services. MatrixPort~\cite{MatrixPort} offers "watch currency rise" OTC options, while Babel Finance~\cite{BabelFinance} provides a ``sharkfin'' capital-protected income management product based on barrier options.

\smallskip
\noindent\textbf{Asset Management.} 
DeFi asset management combines functions of digital assets, oracles, lending, and more DeFi apps to achieve asset management, portfolio management, and risk management. It allows investors to delegate investment decisions to third parties while maintaining trustless functionality. Smart contracts handle investments, trades, and portfolio adjustments based on investor requirements. DeFi asset management offers low start-up costs, and quick set-up times, and enables anyone to become a fund manager or investor.

DeFi asset management can be categorized as active or passive. Active asset management involves a professional team making investment decisions and trades, for example, Enzyme~\cite{Enzyme2023}'s managers or DAO members and Babylon Finance~\cite{Babylon2022}'s community governance. Passive asset management such as Set~\cite{feng2019set} and Index Coop~\cite{indexcoop2020whitepaper}, on the other hand, allows users to create their own indices, structured products, and more in the form of smart contracts. Integrated platforms combine active and passive ways, offering quantitative analysis with machine learning, such as SW DAO~\cite{swdao}, Kava DeFi Platform~\cite{kava2022whitepaper}, and DAOventures~\cite{DAOventures2022}. 

\smallskip
\noindent\textbf{Insurance.} 
DeFi insurance has the same working aspects as traditional insurance, including creation, purchase, and claim of insurance. The differences between DeFi insurance and traditional are that DeFi insurance enables all users to create their own insurance content as can be seen in Etherisc~\cite{cousaert2022token} and turns the decision on insurance claims into a transparent and verifiable process achieved through implementing shared pool models(e.g. Nexus Mutual~\cite{nexus2020}), social proof endorsement(e.g. VouchForMe~\cite{vouch2018}) or prediction markets(e.g. Augur~\cite{peterson2015augur}) and financial derivatives(e.g. oTokens~\cite{koticha2019opyn}).

Insurance is one of the most widely studied applications in DeFi derivatives. The review literature has discussed the potential~\cite{lamberti2018blockchains}  and risk~\cite{singer2019can} of blockchain technology in the insurance industry and the possible application for the entire insurance process~\cite{raikwar2018blockchain}. Various solutions have also been proposed including the construction of the entire framework~\cite{lepoint2018blockcis} and the enhancement of efficiency~\cite{sayegh2019blockchain}, verifiability~\cite{zhang2022identifying}, traceability~\cite{chen2021traceable} and other performance.

\smallskip
\noindent\textbf{Perpetual Contract.} 
DeFi perpetual contracts allow participants to speculate or hedge against the price movements of an underlying asset, similar to leveraged spot trades, but without an expiration date, and use a fund fee mechanism to track the price index of the underlying asset. DeFi perpetual contracts are usually implemented as NFTs by smart contracts and can be traded in DEXs. 
Participants can be incentivized by providing liquidity and receiving rewards.

\smallskip
\noindent\textbf{Prediction Market.} 
Prediction markets involve the creation, trading, and settlement based on real-world event outcomes using smart contracts. Participants are motivated by profit-sharing for accurate predictions and liquidity rewards. The revenue in prediction markets is directly impacted by event outcomes, making it a significant incentive. To determine event outcomes, prediction markets use incentive mechanisms like reward and punishment or oracles providing real-world data. Augur~\cite{peterson2015augur}, for example, incentivizes accurate reporting through a dispute mechanism where the winner receives the loser's staked tokens. Omen Prediction Market~\cite{Omen} introduced Reality.eth, a decentralized oracle challenging previous user results to approach the truth.

\section{Technical Security Risks}
\label{sec-technical}
We identified three types of technical security risks based on DeFi architectural design (Table~\ref{table:Tech-attack}): infrastructure layer risk, protocol layer risk, and application layer attacks.


\subsection{DeFi Infrastructural Layer}



\noindent\textbf{Risks in Network Communication.} 
DeFi relies on network protocols like TCP/IP, which directly impact the security of networks. Attackers can exploit vulnerabilities, manipulate messages, or control network service providers, posing risks to the security of transactions. Denial of Service (DoS) attacks pose a threat where attackers may leverage network congestion to flood the system with invalid transactions or consume excessive bandwidth and computing resources. Additionally, node transparency risks such as Eclipse attacks~\cite{heilman2015eclipse} and Sybil attacks~\cite{10.1007/3-540-45748-8_24} and centralized control by a few entities in 51\% attacks~\cite{saad2020exploring} can undermine trust, security, and stability. Researchers have proposed various approaches to analyze network security, including attack graph analysis~\cite{sheyner2002automated,wang2007toward} and mathematical models quantifying parameters like risk, vulnerability, and threat~\cite{khan2010cyber}.

\smallskip
\noindent\textbf{Risks in Consensus Algorithm.} 
Consensus algorithms enable nodes to reach agreement on tasks such as transaction ordering, block generation, and data validation. Nodes are incentivized with block rewards and transaction fees. However, this decision-making power introduces uncertainty regarding the transactions included in a block, which can be exploited by attackers through Miner Extractable Value (MEV)~\cite{qin2022quantifying}. While MEV can have legitimate uses, such as ensuring timely liquidation in lending protocols, facilitating accurate price formation, and arbitraging in DEX, it also creates problems for users. MEV can result in advantageous forks over the main chain~\cite{perez2021smart,zhang2023time}. Attackers utilize MEV for front-running~\cite{daian2020flash,wang2022cyclic} or sandwich attacks~\cite{zhou2021just}, compromising fairness~\cite{li2023transaction} and colluding with nodes for profit.

\noindent\textbf{\textit{Forks.}}
A fork occurs when the main chain splits into two separate chains. The forked chain may have different security and stability, making it more susceptible to new vulnerabilities and attacks. This can disrupt the compatibility of smart contracts on both chains, requiring redevelopment and migration. In DeFi, chain forks can fragment markets and reduce liquidity. Users may lose funds by mistakenly operating on the wrong chain. Attackers can exploit forks to gain unearned rewards by overtaking and overwriting the main chain.

\begin{figure}   
    \centering
		\subfloat[Front-running Attack]{
		\includegraphics[width=\linewidth]{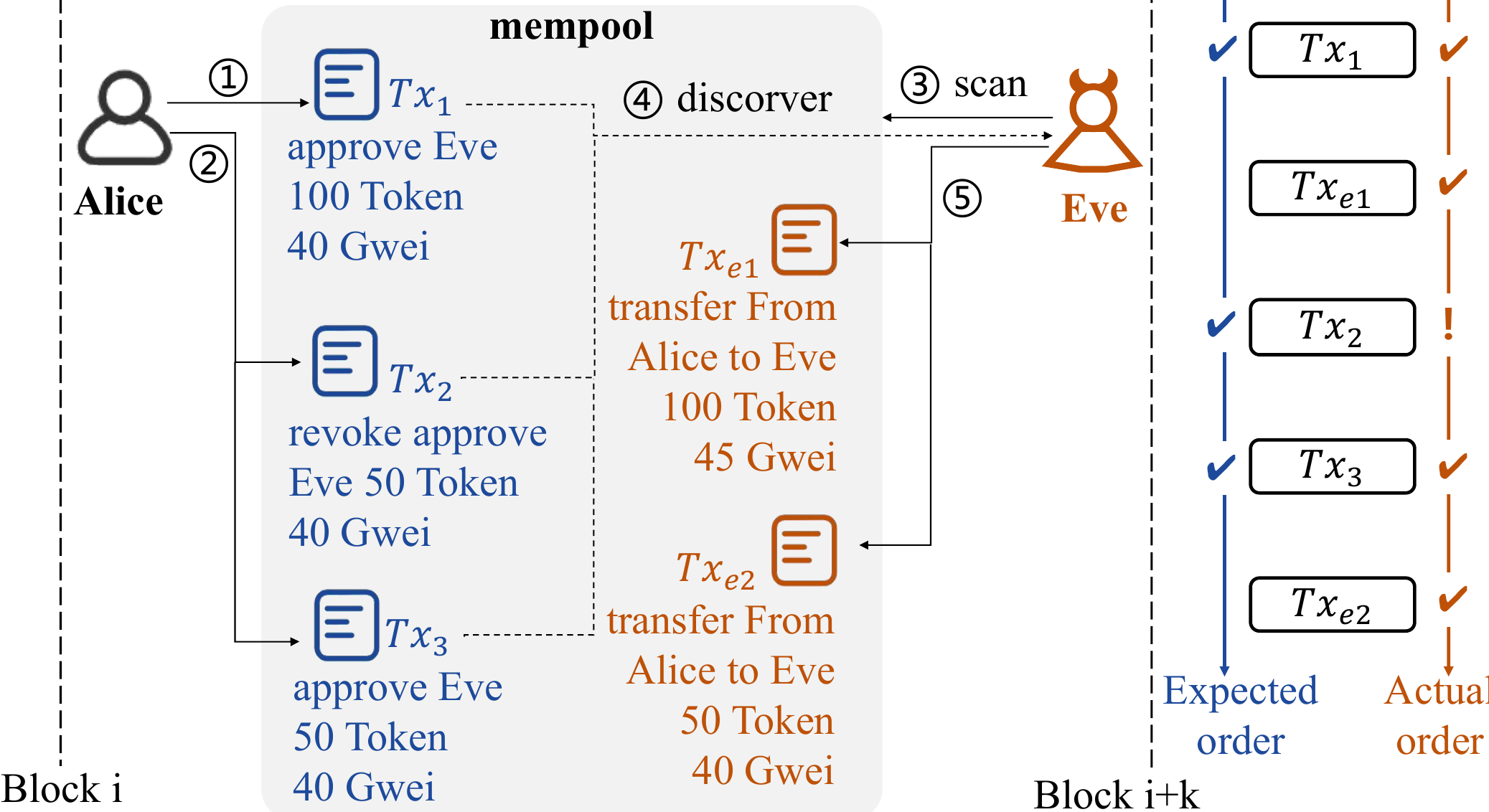}
		\label{fig:front}}
 \quad
		\subfloat[Sandwich Attack]{
		\includegraphics[width=\linewidth]{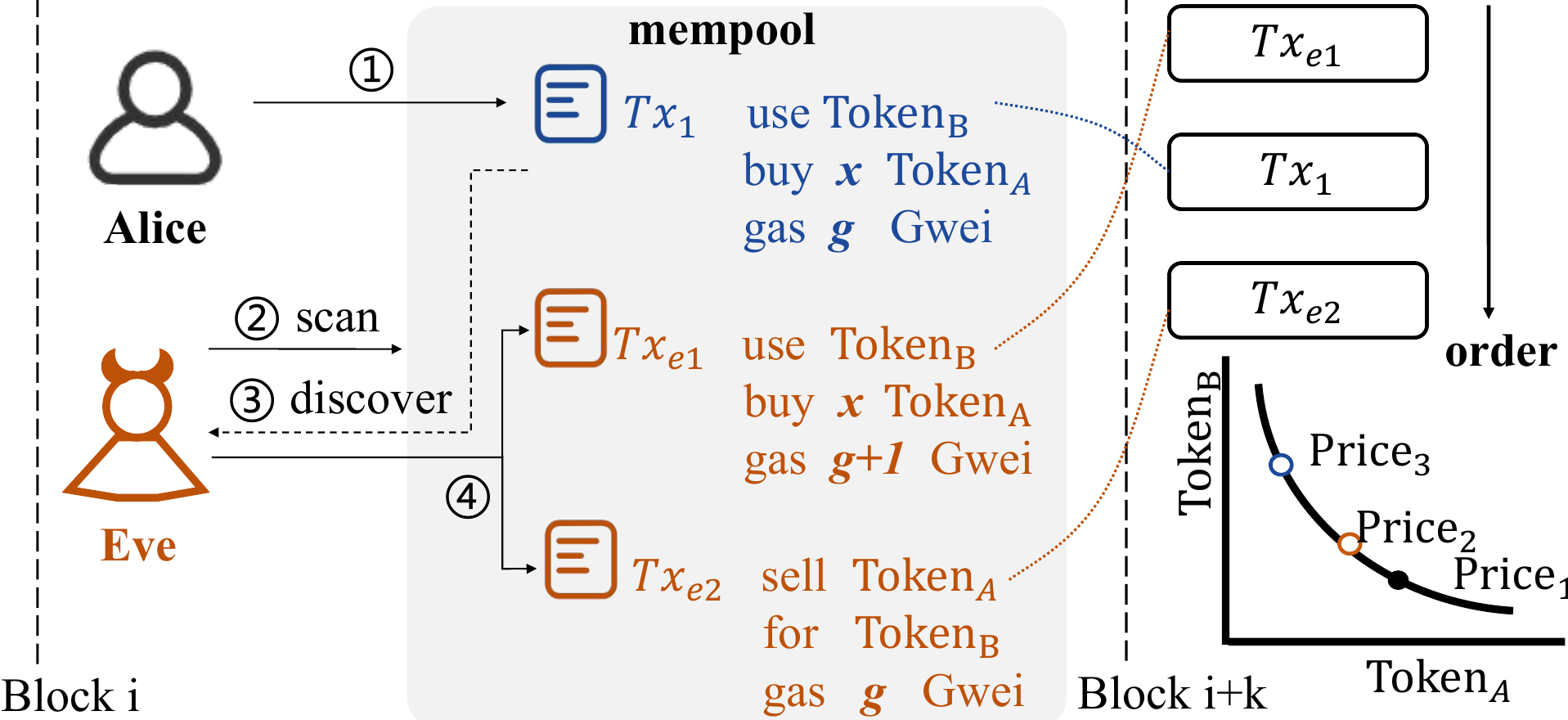}
		\label{fig:sandwich}}
    \caption{Front-running and Sandwich Attack}
\vspace{-0.2in}
\end{figure}

\noindent\textbf{\textit{Front-running.}}
Front-running attacks (cf. Figure~\ref{fig:front}) occur when an attacker predicts or monitors a user's transactions and submits their own transactions with higher priority, blocking others and altering outcomes for additional profit~\cite{torres2021frontrunner,eskandari2020sok}. They exploit blockchain transparency and transaction latency. 
The bZx lending platform suffered a front-running attack in Feb. 2020, where attackers borrowed assets and sold them at manipulated prices, earning significant profits. Mitigation front-running risks solutions include lightning networks for off-chain transactions, batch order processing to narrow the window and raise costs for attackers, sealed transactions to prevent eavesdropping, and fee market efficiency improvement to reduce MEV and front-running profitability~\cite{momeni2022fairblock,zhang2022frontrunning}. FaaS like Flashbots enables traders to directly send transactions to miners, aiming to reduce front-running risks and give users more control. However, Weintraub et al.~\cite{weintraub2022flash} found that over 80\% of Ethereum's MEV occurs through Flashbots, raising questions about the feasibility of FaaS and potential competitive concerns for other participants.

\smallskip
\noindent\textbf{\textit{Sandwich Attack.}}
A sandwich attack~\cite{zhou2021just} (cf. Figure~\ref{fig:sandwich}) is an exploitation tactic where an attacker executes counterparty trades before and after a target trade to profit from price discrepancies and illiquidity. The attacker manipulates the price by squeezing the low-cost trade between the target trade~\cite{wang2022impact}. Attackers monitor DEX order books and trading activity to identify profitable opportunities and swiftly submit counterparty trades for additional revenue. The main difference between sandwich attacks and front-running attacks is the timing of target transaction execution and their respective targets. Sandwich attacks impact prices by executing counterparty trades simultaneously, causing unfair trading losses for the target trader. Front-running attacks gain an advantage by submitting trades before execution, resulting in unfair trading costs for other traders. Although confirming specific sandwich attacks can be challenging, there have been reports of numerous DeFi sandwich attacks exploiting illiquidity, price slippage, and execution delays on DEXs for additional profit.

\subsection{DeFi Protocol Layer}

Smart contracts are vital for implementing and securing DeFi functions. However, they are vulnerable to common vulnerabilities, which have been recognized by academia~\cite{praitheeshan2019security} and industry~\cite{wan2021smart}.  
Beyond coding, improper protocol design can also introduce security risks.

\smallskip
\noindent\textbf{Risks in Writing Smart Contracts.} 
Coding errors such as arithmetic errors, conversion errors, inconsistent access control, and functional reentry are some representative vulnerabilities in smart contracts~\cite{chen2020survey,atzei2017survey}.

\noindent\textbf{\textit{Reentry.}}
A reentry attack is a significant threat to smart contract security. Attackers exploit this vulnerability by repeatedly executing a specific contract function and invoking malicious contracts during each execution~\cite{wang2021mar}. The attacker deploys a malicious contract with callable functions into the target contract and re-invokes it multiple times by calling a function of the target contract. This attack allows unauthorized access to contract funds, modifies the contract status, or performs other malicious actions. The DAO, a community-based investment and fund allocation platform, experienced a reentry attack in 2016. The attacker successfully steals millions of Ether by repeatedly calling the withdrawal function through a malicious contract. To prevent reentry attacks, the Ethereum community has implemented improvements such as a "backward transfer" mode, modifiers to restrict external contract calls, locking mechanisms, status markers, state variables, and lock flags to track and prevent re-calls of functions.

\noindent\textbf{\textit{Overflow.}} 
Overflow is common in smart contracts and can result in unexpected money transfers, contract lockouts, or DoS. These vulnerabilities include integer overflow, array overflow, and memory overflow. Attackers exploit integer overflow to alter contract states or transfer funds. Array overflow allows attackers to access other data in a contract's memory, leading to data tampering. Memory overflow can cause contract execution failure. In 2018, attackers exploited an integer overflow vulnerability in BeautyChain, an Ethereum-based platform, resulting in the theft of approximately \$3 million in cryptocurrency. Similarly, Meerkat Finance, a BSC-based lending protocol, suffered a loss of \$31 million in Mar. 2021 due to an overflow vulnerability. To prevent such vulnerabilities, developers should focus on boundary checking during coding and conduct thorough code reviews.

\noindent\textbf{\textit{Random Numbers Misuse.}} 
Misuse of random numbers in smart contracts can lead to security and fairness issues~\cite{he2020smart}\cite{so2021smartest}. Attackers exploit this vulnerability to predict or manipulate outcomes, gaining unfair advantages. In 2018, hackers manipulated the random number generator of the EOSPlay gambling contract on the EOS blockchain, receiving significant rewards. Insecure random numbers in functions like key generation compromise encryption algorithms. To prevent misuse, developers should carefully assess the need for random numbers. Verifiable generators like Blockchain-based Random Number Generators (BRNGs) and protocols like Distributed Random Number Generation (DRNG) can enhance security. Security audits and code reviews are also essential.

\smallskip
\noindent\textbf{Risks in Updating Smart Contracts.} 
Smart contract updates pose potential issues and security risks such as contract misbehavior, funds loss, contract unavailability, or reduced security. Incompatibility between new and previous versions can introduce vulnerabilities, hinder data migration, or disrupt contract dependencies. Incorrect configuration parameters or tampering can lead to contract failures or unexpected outcomes~\cite{zhang2020txspector}. New permission mechanisms or access control rules may result in incorrect or overly permissive configurations, enabling unauthorized actions. Mismanagement of multiple contract versions can lead to inconsistencies. For instance, in April 2021, Uranium Finance on BSC suffered an attack due to neglected parameter changes during a contract upgrade. To prevent such issues, developers should plan, test, audit upgrades, establish monitoring, and rollback mechanisms to detect and mitigate problems promptly.

\smallskip
\noindent\textbf{Risks in Design of Protocols.} 
Alongside code vulnerabilities, security risks can arise from inadequate protocol design, including logical vulnerabilities, flawed economic models, insufficient risk management, and inappropriate authorization. Complex algorithms or models may overlook specific scenarios, impeding proper functionality. Economic models with inflationary, deflationary, or unfair revenue sharing may lead to revenue loss or instability. Insufficient risk management measures hinder responses to adverse events and risk mitigation. In Jun. 2021, Iron Finance faced a crisis due to its economic model when its governing token TITAN's price collapsed. Massive selling of both TITAN and its stablecoin IRON triggered a mechanism that minted more TITAN as IRON's price dropped, intensifying the price drop and causing a death spiral. Furthermore, flawed designs may grant administrators undue control, enabling manipulation or Rug Pull. The Compounder Finance team misused administrator privileges to replace audited contracts with malicious ones, resulting in the misappropriation of user funds.

\begin{table*}[!ht]
    \centering
    \caption{Concerns and Solutions in DeFi Applications and Open Research Challenges}
    \label{table:Tech-attack}
    \resizebox{0.99\linewidth}{!}{
    \begin{tabular}{clcllllllcccc}
    \toprule
        & \multicolumn{2}{c}{}
        & \multicolumn{2}{c}{}
        & \multicolumn{3}{c}{\textbf{Incidents}}
        & ~ & \multicolumn{4}{c}{\textbf{Research Perspective}}
        \\
        \cmidrule(lr){6-8}
        \cmidrule(lr){10-13}

         \multirow{23}{*}{\rotatebox{90}{\textbf{Technical Security Risk}}}
        & \multicolumn{2}{c}{\textbf{Affected Layers}}
        & \multicolumn{2}{c}{\textbf{Attacks}}
        & Time
        & Application
        & Loss
        & \textbf{Solutions} 
        & Technology
        & Sociology
        & Economy
        & Ecology
        \\
        \cmidrule{2-13} 
        
        & \multirow{8}{*}{\rotatebox{90}{\textbf{\makecell{Infrastructural \\Layer}}}}
        & \multirow{4}{*}{\makecell{Network\\ Communication}}
        &  \multicolumn{2}{l}{\cellcolor{pink!50} Dos (DDoS)~\cite{mirkin2020bdos}\cite{9881505}}
        & \textit{2020/05}
        & \textit{Youbi}
        & \textit{N/A}
        & \cite{9881505}\cite{10.1007/978-3-031-06156-1_19}
        & \cmark
        & & &
        \\

        &
        & 
        & \multicolumn{2}{l}{\cellcolor{pink!50} Eclipse Attack~\cite{heilman2015eclipse}\cite{wust2016ethereum}}
        & \textit{-}
        & \textit{-}
        & \textit{-}
        & \cite{marcus2018low}
        & \cmark
        & & &
        \\
        
        &
        &
        & \multicolumn{2}{l}{\cellcolor{pink!50} Sybil Attack~\cite{10.1007/3-540-45748-8_24}}
        & \textit{Various}
        & \textit{Arbitrum}
        & \textit{253M ARB}
        & \cite{nasrulin2022meritrank} 
        & \cmark
        & & &
        \\
        
        & 
        &
        & \multicolumn{2}{l}{\cellcolor{pink!50} 51\% Attack~\cite{saad2020exploring}\cite{moroz2020double}}
        & \textit{2020/04}
        & \textit{PegNet}
        & \textit{0.6M USD}
        & \cite{bastiaan2015preventing} 
        & \cmark
        & & &
        \\
        \cmidrule(lr){3-9}

        &
        & \multirow{4}{*}{\makecell{Consensus \\Algorithm}}
        &  \multicolumn{2}{l}{\cellcolor{pink!50}  Fork~\cite{9833734}\cite{daian2020flash}\cite{zhou2021just}}
        & \textit{-}
        & \textit{-}
        & \textit{-}
        & \cite{9833734}
        & \cmark
        & & \cmark &
        \\

        & & & \multirow{3}{*}{\rotatebox{90}{\textbf{MEV}}}
        & \cellcolor{pink!50}  Front-running Attack~\cite{eskandari2020sok}\cite{torres2021frontrunner}
        & \textit{2021/03}
        & \textit{DODO}
        & \textit{0.7M USD}
        & \cite{momeni2022fairblock}\cite{zhang2022frontrunning} 
        & \cmark
        & & \cmark &
        \\

        & & &
        & \cellcolor{pink!50}  Sandwich Attack ~\cite{wang2022impact}\cite{zhou2021just}
        & \textit{2021/10}
        & \textit{Alpha Homora V2}
        & \textit{40.93 ETH}
        & \cite{zust2021analyzing}\cite{heimbach2022eliminating}
        & \cmark
        & & \cmark &
        \\

        & & &
        & \cellcolor{pink!50}  Arbitrage Attack~\cite{wang2022arbitrage}
        & \textit{2021/01}
        & \textit{Saddle Finance}
        & \textit{8 BTC}
        & \cite{babel2023clockwork}
        & \cmark
        & & \cmark &
        \\
        \cmidrule{2-9} 

        & \multirow{4}{*}{\rotatebox{90}{\textbf{\makecell{Protocol\\ Layer}}}}
        & \multirow{3}{*}{Smart Contract}
        & \multicolumn{2}{l}{\cellcolor{pink!50} Reentry~\cite{pretre2005attacks}\cite{wang2021mar}}
        & \textit{2016/06}
        & \textit{The DAO}
        & \textit{3.6M ETH}
        & \cite{wang2021mar}
        & \cmark
        & & &
        \\

        & &
        & \multicolumn{2}{l}{\cellcolor{pink!50} Overflow\cite{li2022survey}\cite{oosthoek2021flash}}
        & \textit{2018/07}
        & \textit{Bancor}
        & \textit{1.2M USD}
        & \cite{huang2021hunting}\cite{gao2019smartembed}
        & \cmark
        & & &
        \\

        & &
        & \multicolumn{2}{l}{\cellcolor{pink!50} Misuse of Random Number~\cite{he2020smart}\cite{so2021smartest}}
        & \textit{2021/07}
        & \textit{AnySwap}
        & \textit{8M USD}
        & \cite{gao2019smartembed}
        & \cmark
        & & &
        \\
        \cmidrule(lr){3-9}
        
        & & Protocol
        & \multicolumn{2}{l}{\cellcolor{pink!50} Rug Pull\cite{dos2022new}}
        & \textit{2021/03}
        & \textit{Meerkat Finance}
        & \textit{20M USD}
        & \cite{mazorra2022not}
        & \cmark
        & \cmark &  &
        \\
        \cmidrule{2-9} 
        
        & \multirow{9}{*}{\rotatebox{90}{\textbf{\makecell{Application\\ Layer}}}}
        & \multirow{7}{*}{Bridge}
        & \multicolumn{2}{l}{\cellcolor{pink!50} Double Spend Attack\cite{lee2023sok}}
        & \textit{2020/02}
        & \textit{DForce}
        & \textit{2.5M USD}
        & \cite{sai2019disincentivizing}
        & \cmark
        &  &  &
        \\

        & &
        & \multicolumn{2}{l}{\cellcolor{pink!50} False Proof Attack\cite{lee2023sok}\cite{herlihy2019cross}}
        & \textit{2022/02}
        & \textit{Wormhole}
        & \textit{1.2M ETH}
        & \cite{herlihy2019cross}
        & \cmark
        &  &  &
        \\

        & &
        & \multicolumn{2}{l}{\cellcolor{pink!50} Replay Attack~\cite{sonnino2020replay}}
        & \textit{2022/09}
        & \textit{OmniBridge}
        & \textit{2M ETHW}
        & \cite{han2023survey}\cite{duan2023attacks}
        & \cmark
        &  &  &
        \\

        & &
        & \multicolumn{2}{l}{\cellcolor{pink!50} Inter-Chain Route Hijacking~\cite{careem2020reputation}}
        & \textit{-}
        & \textit{-}
        & \textit{-}
        & \cite{lv2022attack}\cite{duan2023attacks}
        & \cmark
        &  &  &
        \\

        & &
        & \multicolumn{2}{l}{\cellcolor{pink!50} Wormhole Attack~\cite{malavolta2018anonymous}}
        & \textit{-}
        & \textit{-}
        & \textit{-}
        & \cite{sun2022decentralized}
        & \cmark
        &  &  &
        \\

        & &
        & \multicolumn{2}{l}{\cellcolor{pink!50} Cross-chain Price Manipulation~\cite{harris2023cross}}
        & \textit{2023/08}
        & \textit{Neutra Finance}
        & \textit{23.5 ETH}
        & \cite{harris2023cross}\cite{wu2021defiranger}
        & \cmark
        & & \cmark &
        \\
        \cmidrule(lr){3-9}

        & & Auxiliary Tools
        & \multicolumn{2}{l}{\cellcolor{pink!50} Oracle Manipulation~\cite{su2021evil}\cite{werner2022sok}\cite{mackinga2022twap}}
        & \textit{2023/06}
        & \textit{Themis Protocol}
        & \textit{0.37M USD}
        & \cite{eskandari2021sok}
        & \cmark
        & & \cmark &
        \\
        \cmidrule(lr){3-9}

        & &Usage Method
        & \multicolumn{2}{l}{\cellcolor{pink!50} Phishing Attack~\cite{malavolta2018anonymous}\cite{winter2021s}}
        & \textit{2022/12}
        & \textit{Bitkeep}
        & \textit{8M USD}
        & \cite{li2021self}\cite{wang2022tsgn}
        & \cmark & \cmark &  & \cmark
        \\
        
        \cmidrule(lr){1-9}
        \multirow{8}{*}{\rotatebox{90}{\textbf{Economic Risk}}} & & & & & & \multicolumn{3}{c}{\textbf{Real Collapse/Incidents}}   \\
        \cmidrule(lr){6-9}
        
        &\multirow{7}{*}{\rotatebox{90}{\textbf{\makecell{Infrastructural \\ + Protocol  \\ + Applicatoin \\Layers  }}}} & &  \multicolumn{2}{l}{\cellcolor{pink!50} Liquidity Depletion} & \textit{2019/09} & \textit{Compound} & - & \cite{illiquidity2019alethio}\cite{aramonte2021defi} 
        &  &  & \cmark &\\

        & &  & \multicolumn{2}{l}{\cellcolor{pink!50} Governance Risk} & \textit{2022/04} & \textit{Beanstalk} & \textit{77M USD} & \cite{beanstalk2022bean}\cite{mondoh2022decentralised} 
        & \cmark
        & \cmark & \cmark  & \cmark\\

        & &  & \multicolumn{2}{l}{\cellcolor{pink!50} Market Manipulation}& \textit{2022/10} & \textit{Mango} & \textit{114M USD} & \cite{spoof2021reiff}\cite{auer2022miners}
        & \cmark & \cmark & \cmark  & \cmark\\

        &  & & \multicolumn{2}{l}{\cellcolor{pink!50} Flashloan} & \textit{various} & \textit{various} & \textit{various} & \cite{daian2020flash}\cite{qin2021attacking}
        & \cmark &  & \cmark  & \\

        & &  &  \multicolumn{2}{l}{\cellcolor{pink!50} Death Spiral} &  \textit{2022/05} & \textit{Luna-UST Collapse} & \textit{60B USD} & \cite{kerr2023cryptocurrency}\cite{fu2023rational}
        &  &  & \cmark & \\

        & &  &  \multicolumn{2}{l}{\cellcolor{pink!50} Ponzi and Fraud} & \textit{2022/11} & \textit{FTX Collapse} & \textit{32B USD} & \cite{fu2022ftx}
        & & \cmark & \cmark  & \cmark\\

        & &   & \multicolumn{2}{l}{\cellcolor{pink!50} CeFi Impact} & \textit{2023/03} & \textit{SVB Collapse} & \textit{23B USD} & \cite{yousaf2023impact}\cite{wang2023cryptocurrency}
        & &  & \cmark  & \\

        \bottomrule 
    \end{tabular} }
    \vspace{-0.1in}
\end{table*}

\subsection{DeFi Application Layer}

The security risks of DeFi extend beyond the system's internal workings to include external attacks towards asset bridges, irregular services provided by auxiliary applications like oracles, and users' misconceptions about smart contracts.

\smallskip
\noindent\textbf{Risks in Cross-chain.} 
Cross-chain attacks exploit the mechanism of cross-chain transactions, posing risks to the security and stability of cross-chain DeFi apps. These attacks can lead to asset loss, transaction delays, and information tampering. There are two main types of cross-chain attacks: native-chain attacks and inter-chain attacks. Native-chain attacks include double-spend attacks, false proof attacks, vulnerability exploits, reverse transaction attacks, and replay attacks~\cite{sonnino2020replay}. Inter-chain attacks encompass relay blocking and inter-chain route hijacking~\cite{careem2020reputation}. Payment channels may also be vulnerable to wormhole attacks, where intermediate node fees can be stolen~\cite{malavolta2018anonymous}. DeFi cross-chain applications face unique security risks. Cross-chain smart contracts in DeFi apps are exposed to vulnerabilities in their own code and the calling relationship between contracts. Price manipulation attacks and repeated borrowing and lending attacks are examples of cross-chain attacks faced by DeFi apps, as seen in the case of the attack on PancakeSwap in Apr. 2021.

\smallskip
\noindent\textbf{Risks in Auxiliary Tools.} 
Auxiliary services are entities that promote efficiency but are external to the system. 

\noindent\textbf{\textit{Oracle Manipulate.}} 
Oracle manipulation by hackers involves providing false data to smart contracts, leading to improper benefits or disruption of normal operations~\cite{su2021evil}. This manipulation can result in negative consequences, including stablecoin unanchoring, malicious carry trades, forced liquidation, and depleted protocol liquidity. Attackers target data sources by attacking API interfaces or tampering with supply chains. They provide false or inaccurate data, modify prices or offer incorrect market information. To prevent oracle manipulation, developers must prioritize secure and tamper-resistant oracles, along with incentivizing their usage. Ensuring the quality of connected markets is also crucial.

\smallskip
\noindent\textbf{Risks of Ignorance.} 
Users' limited understanding of smart contracts and their associated security risks can lead to unforeseen circumstances~\cite{coingeco2020survey}. Moreover, the shortage of security awareness makes them susceptible to phishing attacks, resulting in personal information leaks and fund theft~\cite{schar2021decentralized}. Phishing attacks in DeFi involve impersonating legitimate entities or creating deceptive environments like fake DeFi platforms or sending fraudulent notifications to trick users into revealing sensitive information, private keys, or login credentials. For example, in Dec. 2021, Badger DAO suffered a \$120 million loss due to a phishing attack involving malicious wallet requests. Similarly, in 2021, attackers stole assets by sharing fraudulent links on social media, leading users to a fake Uniswap website.

\section{Economic Security Risks}
\label{sec-economic}
DeFi economic risks stem from rational players' actions within the ecosystem, rather than traditional vulnerabilities.

\smallskip
\noindent\textbf{Liquidity Depletion.} 
DeFi liquidity depletion risk occurs when there is a shortage of market liquidity, causing transaction delays, price fluctuations, and market instability. Insufficient liquidity can result from adverse market conditions, increased risk sentiment, or rapid fund withdrawals. Factors like market fluctuations, falling collateral prices, or manipulation can also contribute to liquidity depletion. The Black Thursday event in Mar. 2020, involving MakerDAO, exemplifies the consequences of liquidity depletion~\cite{wang2022empirical}. To mitigate this risk, DeFi platforms should attract diverse liquidity providers and reduce dependency on specific sources. Incentives can be implemented to attract and retain liquidity providers~\cite{heimbach2021behavior,heimbach2022risks}. Additionally, DeFi applications should develop risk management strategies and contingency plans to address liquidity depletion scenarios.


\noindent\textbf{Flashloan Attack.} 
Flash loan security risks involve vulnerabilities, contracts, and attack risks associated with flash loans~\cite{daian2020flash,qin2021attacking}. We observed that such attacks occur frequently and can be categorized into code vulnerabilities, bid arbitrage, and price manipulation. Attackers exploit flash loans to execute sophisticated strategies that exploit smart contract vulnerabilities. Bid arbitrage manipulates transaction sequences to capitalize on arbitrage opportunities using borrowed funds. Price manipulation involves using flash loan funds for large-scale trading, influencing prices. Defending against flash loan attacks requires managing protocol security, auditing contracts, and implementing measures like transaction order restrictions, delays, or time windows.

\smallskip
\noindent\textbf{Market Manipulation.} 
Market manipulation artificially influences asset prices to profit. Illiquid assets pose higher risks to underlying financial products. Manipulative strategies include spoofing~\cite{cernera2022token}, ramping~\cite{xu2019anatomy}, bear raids, cross-market manipulation, and oracle manipulation~\cite{eskandari2021sok}, which can manipulate segments or the entire market. Market manipulation has resulted in various negative impacts on the DeFi ecosystem, such as bad debts due to failure of timely liquidation, losses for liquidity providers due to false price-based payouts in synthetic assets, and depeg for algorithm stablecoins~\cite{fu2023rational}.

Distinguishing normal fluctuations from manipulation in DeFi is challenging due to anonymity, trading freedom, and regulatory gaps.   Anonymous transactions hinder accurate tracking of participant behavior. Trading freedom and liquidity provision enable price influence through large-scale transactions or exploiting limited market depth. Regulatory gaps and the lack of mechanisms like KYC requirements hamper monitoring manipulative behavior. To reduce market manipulation risks in the DeFi market, it is necessary to strengthen regulatory compliance and enhance investor education. Monitoring tools and algorithms can detect abnormal trading patterns and manipulative behavior in a timely manner. Strengthening investor education can increase awareness of market risks, encouraging cautious participation.

\smallskip
\noindent\textbf{Governance Risk.} 
Governance and incentives can drive choices that benefit DeFi apps. However, inadequate incentives may lead token holders to prioritize external gains, potentially harming the system. Immediate governance updates can be vulnerable if malicious contract code is executed using acquired governance tokens. The Beanstalk protocol faced governance risks when an attacker accumulated tokens and proposed a malicious proposal to divert funds. In Ethereum 2.0 (post-Merge), validators face censorship pressure due to The Office of Foreign Assets Control of the US Department of the Treasury(US OFAC) sanctions on Tornado Cash~\cite{wang2023blockchain,wahrstatter2023blockchain}.

\smallskip
\noindent\textbf{Death Spiral and Instability.} 
Stablecoins aim to maintain a consistent value by being pegged to a fiat currency. A "death spiral" refers to a scenario where a stablecoin rapidly and uncontrollably loses its value. This decline can trigger a compounding effect, intensifying the downward spiral and eroding confidence in the stablecoin, ultimately culminating in a self-perpetuating cycle of value deterioration. The concept of a death spiral is particularly relevant in crypto stablecoins, especially algorithmic ones. An example is the deppeg of UST, which led to the collapse of Luna's value in 2022\cite{kerr2023cryptocurrency,fu2023rational}. This event marked the onset of a "crypto winter," a period of prolonged market decline and reduced investor optimism.

\smallskip
\noindent\textbf{Ponzi and Fraud.} 
The Ponzi game is named after Charles Ponzi, who deceived investors with a postage stamp speculation scheme. It involves funding preexisting liabilities by issuing new debt. Ponzi schemes have infiltrated the crypto markets, especially during ICOs, IEOs, IDOs, and similar events. A recent example is the FTX collapse~\cite{fu2022ftx} in 2022, where SVB sold FTT tokens to Alameda, a high-frequency trading company under the same ownership, in an attempt to inflate token prices. Ponzi's collapse extends beyond immediate losses of rug pull with far-reaching implications.

\smallskip
\noindent\textbf{CeFi Impact.} 
The crypto market has become increasingly influenced by macroeconomics, exhibiting a trajectory that parallels traditional stock markets. It is susceptible to the impacts of conventional financial crises. The SVB (Silicon Valley Bank) collapse~\cite{yousaf2023impact,wang2023cryptocurrency} in early 2023 is an illustrative case. SVB, a Centralized Finance (CeFi) bank for high-tech and crypto startups, experienced a crisis that resulted in the loss of deposits. This event eroded confidence in secure crypto asset storage methods and raised concerns about asset safety.

\section{Open Research Challenges}
\label{sec-direction}



\noindent\textbf{Observation.} There exists a time lag between the industry and academia, as industry innovation often precedes academic research. However, the implementation of DeFi in the industry currently is primarily driven by commercial intuition, lacking comprehensive academic research in problem definition, theoretical analysis, mechanism design, and economic models. This gap hinders theoretical innovation and the timely resolution of emerging issues, limiting efficiency and sustainability.

\smallskip
\noindent\colorbox{babypink}{\textbf{Gap.}} The relative lag in academic research, particularly with regard to information security concerns, can result in potential inadequacies in the security assessment of newly emerging DeFi protocols. Accidents and risks that arise during the use of new technologies (emerging financial products) cannot be anticipated, forewarned, or prevented in advance, posing risks to user assets and impeding comprehensive development. 

\smallskip
\noindent\colorbox{cambridgeblue}{\textbf{Revenue.}} Therefore, further research on DeFi requires in-depth exploration of new technologies and models, encompassing different perspectives such as information security, and game theory mechanisms, and conducting systematic evaluations that are aligned with industry practices. To bridge the gap between industry and academia, it is necessary to address the lack of research and validation tools in the academic community. Academic researchers require models, simulation tools, and data analysis capabilities to establish and validate DeFi solutions.   Effective data and analysis tools are needed to collect, organize, and analyze transaction and contract data.   Comprehensive security audit tools are also necessary to evaluate the security of smart contracts and protocols. Joint efforts between academia and the industry can promote the development and improvement of these tools. Academic researchers can focus on developing models and simulation tools, while the industry can provide real-time data and practical experience to support data and analysis tool development.   Collaboration between the academic community and security audit teams can enhance the security of DeFi projects by jointly researching and developing security audit tools.

\smallskip
\noindent\colorbox{babypink}{\textbf{Gap.}} Regarding economic security concerns, the relative lag in academia has resulted in a lack of in-depth analysis and theoretical modeling of economic interactions and mechanism designs in DeFi. Firstly, the industry lacks a proper understanding of participant behavior and incentive mechanisms, leading to product designs that heavily rely on experience and intuition. Secondly, information asymmetry and incomplete information exacerbate the industry's limited awareness and application of existing academic research. DeFi applications lack guidance from economic equilibrium theories, leading to potential risks, instability, manipulation, attacks, and systemic risks. Inappropriate incentive structures and market imbalances undermine the achievement of economic and societal objectives, efficiency, fairness, and sustainability.

\smallskip
\noindent\colorbox{cambridgeblue}{\textbf{Revenue.}}  The academic community should utilize analytical tools such as economic equilibrium and game theory to establish technical and economic models for DeFi applications and conduct research on game mechanism design and analysis. Progress can be made from multiple perspectives of economic modeling, mechanism design, and technical implementation, and combining theoretical analysis, practical solutions, and empirical research.

\smallskip
\noindent\colorbox{babypink}{\textbf{Gap.}} Regardless of the improvement in technology, security, or economics in the field of DeFi, the interdisciplinary nature of DeFi cannot be overlooked. Collaboration between economists, computer scientists, legal experts, and other stakeholders is crucial. As mentioned earlier, the limitations of a single-disciplinary perspective have been highlighted, and it has been pointed out that the core issue lies in the insufficient depth of interdisciplinary collaboration and the shortage of talent with interdisciplinary expertise. 

\smallskip
\noindent\colorbox{cambridgeblue}{\textbf{Revenue.}} To address this issue, on one hand, it is necessary to encourage and facilitate interdisciplinary collaboration among experts from different disciplines through cross-disciplinary projects or platforms. On the other hand, it is important to cultivate talents with multidisciplinary skills required in DeFi research. Specifically, these talents need to possess expertise in areas such as blockchain, network security, code analysis, financial markets, investment analysis, risk management, and financial technology, spanning disciplines like economics, computer science, and cryptography.

In addition to the overall observations in DeFi mentioned above, there are also research gaps and challenges in specific aspects such as technological construction, sociological construction, economic construction, and ecosystem construction. We will elaborate on each of these areas (cf. Table~\ref{tab:openresearch}). 

\begin{table}[!htb]
    \centering
    \caption{Open Research Challenges} \label{tab:openresearch}
     \resizebox{\linewidth}{!}{
    \begin{tabular}{lll}
    \toprule
          \textbf{Direction}
          & \textbf{Open Research Challenge}
         & \textbf{Literature (paper count)}\\
        \midrule
        
        General Tools
        & \cellcolor{pink!50} Definition and Model
        & 283, e.g., \cite{werner2022sok}  \\ 
        \midrule
        
       \multirow{4}{*}{\makecell{DeFi Technical \\ Construction}}
        & \cellcolor{pink!50} Performance
        & 137, \cite{6331994}\cite{10240601}\cite{HaoDing-4349}  \\ 
        
         ~
         & \cellcolor{pink!50} Function Integration Platforms
         & N/A\\ 
         
         ~
         & \cellcolor{pink!50} Contract Audition
         & 600, \cite{JK-3087}\cite{ZhangLi-3119}\cite{MaLiu-3127}   \\ 
         
         ~
         &\cellcolor{pink!50} Incident Detection
         & 308, \cite{YeShen-2850}\cite{Mercy_PraiseBasil_Xavier-2855}\cite{SunXiao-2880}   \\ 
         \midrule
         
       \multirow{2}{*}{\makecell{DeFi Economy \\ Construction}}
         & \cellcolor{pink!50} Sustainable Tokenomics
         & 193, \cite{LetychevskyiPeschanenko-3628}\cite{ZhihongJie-3629}\cite{Lesche2022} \\ 
         
         ~
         & \cellcolor{pink!50} Balanced Incentive
         & 122, \cite{B.J.-102}\cite{GanSaini-668}\cite{JudmayerStifter-2968}  \\ 
        \midrule
        
        \multirow{2}{*}{\makecell{DeFi Sociology\\ Construction}}
        &\cellcolor{pink!50} Privacy
        & 216,  \cite{9680519}\cite{HartmannHasan-4347}   \\ 
        
        ~
         & \cellcolor{pink!50} Compliance
         & 75, \cite{Blemus-2815}\cite{UshidaAngel-2826}\cite{Laguna_De_Paz-2830} \\ 
        \midrule
        
        DeFi Ecology 
        & \cellcolor{pink!50} User Engagement and Education
        & 11, \cite{Reichert-4300}\cite{Saiya-4302}\cite{MedeirosDeligiannidis-2947}\cite{Y.H.-63}  \\ 
        \bottomrule
    \end{tabular}}
\vspace{-0.5em}
\end{table}

\subsection{DeFi Technology Construction}
\textbf{Functionality.} 
There are numerous protocols and functionalities in DeFi. Function integration platforms can provide unified access, simplifying user operations and enhancing user experience. Their goals include user-friendly interfaces, protocol integration, security, and interoperability for seamless asset and data transfers. However, these aspects have not been extensively explored by both the industry and the academic.

Research challenges for function integration platforms encompass staying updated and incorporating the developments and innovations, implementing better risk management practices to adapt to the dynamic nature of the DeFi market, conducting comprehensive security audits and vulnerability fixes, and addressing challenges related to interoperability, stability, and availability that arise from integrating multiple protocols.

\smallskip
\noindent\textbf{Security.}
DeFi security involves analyzing attack and threat models at the network~\cite{sheyner2002automated,wang2007toward,khan2010cyber}, smart contract~\cite{tsankov2018securify} (single contract, multiple contracts, and contract audit~\cite{zhou2022sok}), protocol~\cite{bano2019sok}, and application layers~\cite{amin2019practical,woods2021sok}. 

We found that in-depth research on DeFi network communication security, standardized evaluation, audit methods, and defense strategies is lacking.   Researchers can enhance DeFi security by analyzing network topology, node communication, and protocols like authentication, access control, and secure smart contracts.   Existing smart contract audit tools have limitations in detecting complex attack strategies and advanced vulnerabilities.   Research is needed on tracking contract changes, conducting timely audits, and establishing security standards and best practices.   Moreover, there is limited research on secure DeFi application models and security issues arising from the collaboration between different applications.

\smallskip
\noindent\textbf{Incident Detection and Emergency Response.} 
DeFi acts as an amplifier for information security issues and risks, making disaster recovery and emergency response more urgent. Currently, this is a relatively unexplored research field. Timely incident detection and handling are essential for protecting user assets and the health of DeFi. Detecting incidents based on historical data is commonly done, but real-time monitoring remains understudied. Future research should focus on developing intelligent monitoring systems that analyze data and traffic patterns, using machine learning to identify abnormal activities and risks in advance. Establishing effective incident response mechanisms, as well as post-incident cooperation and asset recovery protocols, are important research gaps.

\subsection{DeFi Sociology Construction}

\textbf{Privacy.} 
DeFi privacy protection aims to safeguard users' personal information, transaction data, and financial flows from unauthorized tracking, monitoring, and access. As the user base expands, privacy concerns in DeFi become more apparent, as existing analysis techniques can de-anonymize pseudonyms and infer user identities from external information on the blockchain~\cite{meiklejohn2013fistful,androulaki2013evaluating}. Existing research explores privacy-enhancing technologies such as zero-knowledge proofs (ZKPs) \cite{wang2023zero}, ring signatures \cite{sun2017ringct}, Trusted Execution Environments (TEEs) for anonymous computation, cryptographic techniques, and mixing schemes for transaction privacy protection, as well as confidential smart contracts for data privacy protection~\cite{li2022sok,li2020accountable}. Initially applied to privacy coins like Zerocoin~\cite{miers2013zerocoin}, zero-knowledge proofs (ZKPs) such as zk-SNARKs~\cite{groth2016size} and Bulletproofs~\cite{bunz2018bulletproofs} have been utilized to enhance transaction privacy on the blockchain. Simpler and more efficient privacy coin schemes, including mixing schemes, have also been proposed. These technologies offer improved privacy protection, data security, anonymous transactions, and smart contract verification in DeFi. However, how these technologies can adapt to the specific and complex application scenarios in DeFi, enhance performance, and achieve scalability, remains an unresolved research question. Additionally, striking a balance between anonymity and traceability to meet regulatory requirements poses significant challenges. Readers interested in exploring privacy-enhancing technologies in DeFi are recommended to refer to relevant Systematization of Knowledge (SoK) papers~\cite{baum2023sok}.


\smallskip
\noindent\textbf{Compliance.} 
Compliance in DeFi is vital for protecting funds for users, ensuring stability for the system, and preventing illegal activities such as money laundering and terrorist financing for regulators.  However, the decentralized and permissionless nature of DeFi conflicts with compliance and regulation. For instance, Tornado Cash faced sanctions by the US OFAC for enabling anonymous transfers used in illicit activities. Balancing these conflicting aspects poses a challenge. Moreover, the rapid growth of DeFi requires the development of an appropriate compliance framework, including cross-border cooperation and leveraging technology for efficient processes.

\subsection{DeFi Economy Construction}



\smallskip
\noindent\textbf{Sustainable Tokenomics.}  
Token issuers should prioritize the design and implementation of a cryptocurrency or token ecosystem that promotes long-term viability, equilibrium, and positive impact. Rather than engaging in deceptive practices, their focus should be on crafting mechanisms that enhance economic stability, while incentivizing productive and sustainable behaviors among token holders, users, and participants. Initiatives should be geared towards creating enduring value for the token, rather than relying on short-term speculative gains. By contributing to a resilient token ecosystem characterized by responsible practices and sustainable growth, token issuers can foster a positive and sustainable impact.

\begin{figure}[!hbt]
    \centering
    \includegraphics[width=1\linewidth]{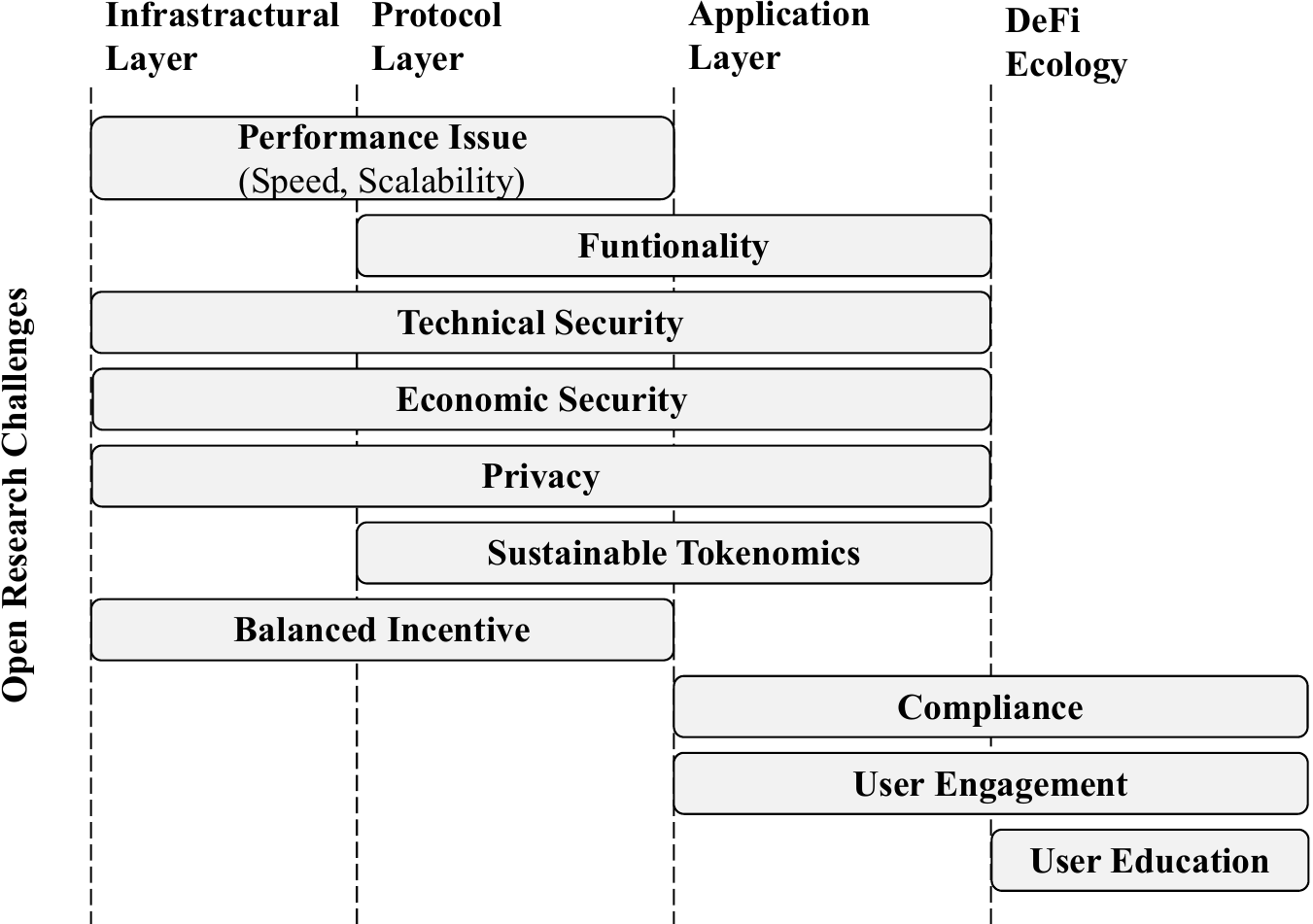}
    \caption{Open Research Challenges}
    \label{fig:openresearch}
\end{figure}

\smallskip
\noindent\textbf{Balanced Incentive.} 
Incentives hold a crucial role for token investors and holders. An effective incentive mechanism should extend benefits to all participants, encompassing both regular users (utilizing mobile clients, light clients, or web browsers) and network contributors (such as miners, validators, and operators). Within the PoW consensus model, the majority of regular users often find it challenging to secure stable rewards, given their limited impact on the network. In contrast, miners wield substantial power, affording them the ability to generate MEV revenues. Meanwhile, in PoS settings, stakeholders have the opportunity to earn a steady income. However, the issuance of tokens without associated costs can lead to potential token issuance abuse. Striking a balance between incentivizing network contributors while preventing token issuance abuse is crucial. This will create a fair ecosystem where rewards align with contributions, promoting healthier and equitable participation from all stakeholders.

\subsection{DeFi Ecology Construction}

\textbf{User Engagement and Education.} 
User engagement and user education are crucial for the DeFi ecosystem's growth. Low user engagement can lead to reduced liquidity, increased transaction costs, and insufficient market depth, while user ignorance can elevate the risk of fraud and attacks. Enhancing usability is paramount in addressing these challenges. The industry should prioritize creating user-friendly interfaces and streamlining processes to attract a broader user base. Additionally, effective incentive mechanisms can encourage user participation and contributions. Comprehensive and easily understandable educational resources, including textbooks, online courses and guides, must be made available. Establishing a supportive community can further bolster user education by promoting collaboration and knowledge sharing, enabling users to assist and support each other on their DeFi journey.

\smallskip
\noindent\textbf{\large Conclusion.} This paper presents a comprehensive literature review and proposes a classification framework based on the complexity of DeFi services. We evaluate DeFi applications and discuss their security from both technical and economic perspectives. Furthermore, we identify research gaps and future directions, exploring relevant research topics. 

\bibliographystyle{unsrt}
\bibliography{bib.bib}

\appendix

\setcounter{figure}{0}
\section{Foundations of DeFi}
\label{sec-foundation}
\subsection{Operational Supports}

\begin{figure}[!htbp]
    \centering
    \includegraphics[width=0.95\linewidth]{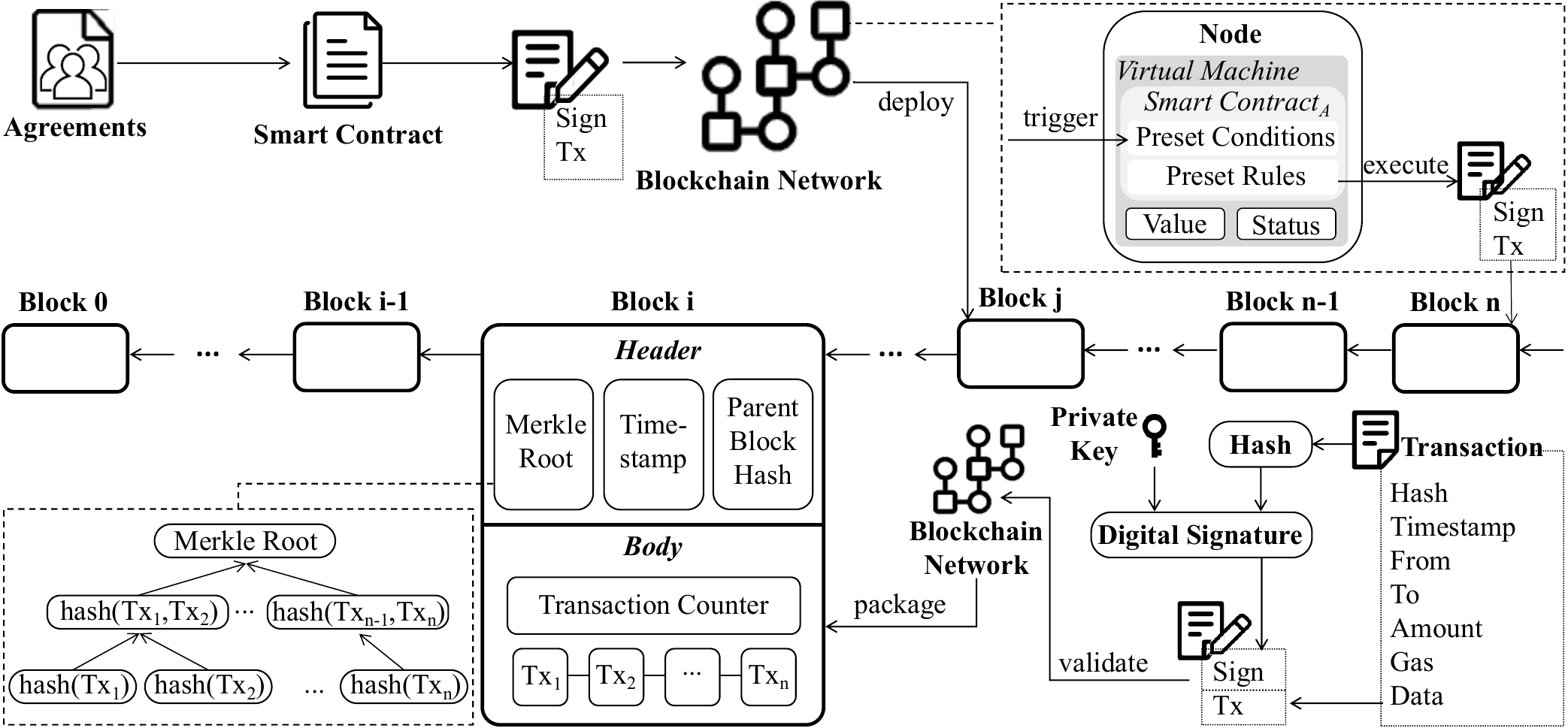}
    \caption{DeFi Foundations}
\label{fig:foundations}
\vspace{-0.1in}
\end{figure}

\noindent\textbf{Transactions.} A transaction is the smallest unit in the blockchain ledger. It includes sender and receiver addresses, the number of coins involved, a unique hash value, a timestamp, transaction/gas fee, block information (block ID of the first recording block), and data payloads for execution (cf. Figure~\ref{fig:foundations}). Interactions with the blockchain are categorized as transfer or contract transactions. Transfer transactions involve simple coin transfers, while contract transactions interact with smart contracts. A transaction sender must be an  Externally Owned Account (EOA), while the receiver can be a smart contract address or an EOA, and the transaction data field contains the required parameters for the contract function.

\smallskip
\noindent\textbf{Block.}  The block is a fundamental unit of data, consisting of header and body. The header contains the previous block's hash, current block's ID, and Merkel root of its content, ensuring a tamper-proof chain. The block body contains transactions. Creating a new block involves propagation and validation across different nodes via consensus algorithms. A newly added block is linked in the current chain.

\smallskip
\noindent\textbf{Chain.}
The chain is a series of blocks linked together using cryptographic hashes (cf. Figure~\ref{fig:foundations}). Each block contains a unique identifier (hash) derived from its data and the previous block's hash. This creates a continuous and tamper-resistant chain of data known as the blockchain (conceptual milestones in 1991~\cite{haber1991time}, 2008~\cite{nakamoto2008bitcoin}, and 2014~\cite{wood2014ethereum}).

\smallskip
\noindent\textbf{Smart contracts.} Smart contract constitutes a crucial element supporting DeFi protocols. Deployed on-chain, it acts as a computerized transaction protocol that transforms traditional contract terms into executable programs, maintaining logical connections between terms as a flow (see Figure~\ref{fig:foundations}). Smart contracts feature automatic execution, instant response, and strict enforcement, and the contracts deployed on them are tamper-proof, minimizing the chance of human intervention.

\smallskip
\noindent\textbf{DApp.} Short for decentralized applications, DApps are constructed on blockchain using smart contracts~\cite{popescu2020decentralized}. Smart contracts can be likened to code-based Lego blocks with automatic execution functions~\cite{katona2021decentralized}. Multiple smart contracts can collaborate to achieve the intricate functionalities required by applications. DApps usually offer user interfaces, streamlining users' interactions with the blockchain. User actions via DApps are recorded on the blockchain as transactions, executed according to pre-written smart contract rules, and verified by blockchain nodes.

\subsection{DeFi Composition}

\begin{figure}[!hbt]
    \centering
    \includegraphics[width=0.95\linewidth]{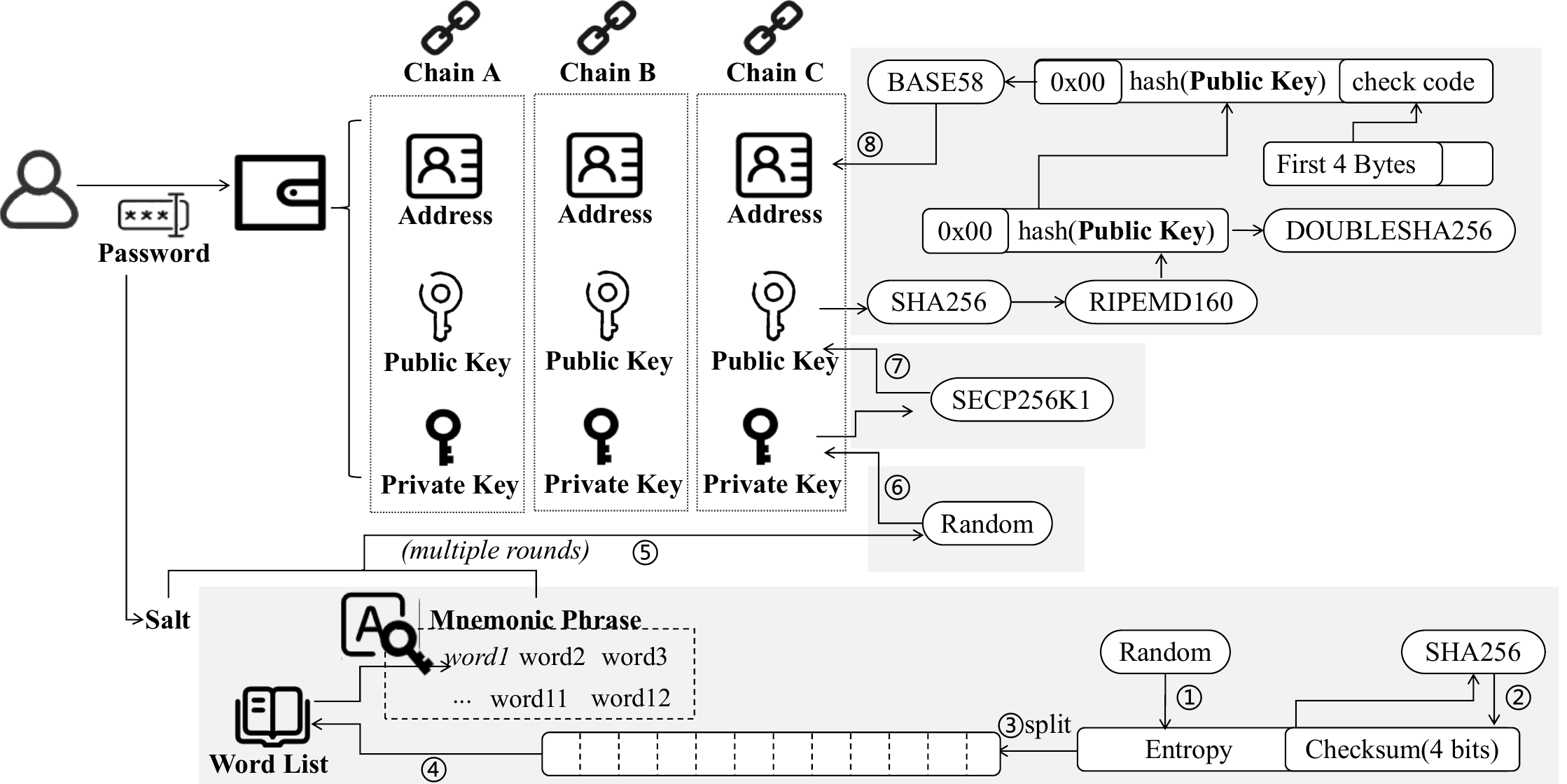}
    \caption{Components of Wallets}
    \label{fig:wallet}
\end{figure}

\noindent\textbf{Wallet.} A user can manage multiple accounts from a single wallet in DeFi. Each account has three components: public key, private key, and address, as shown in Figure~\ref{fig:wallet}. A cryptographic algorithm generates a pair of one-to-one keys when an account is created. The private key generates the digital signature necessary for proving ownership of assets, which can be verified by the corresponding public key. An address, generated from the public key by a one-way hash function, is to DeFi what an account is to traditional finance, symbolizing a user's on-chain identity. Since private keys are difficult to remember, the wallet developers have set up mnemonics as double insurance policy to help users memorize complex private keys. A mnemonic can be understood as a simplified version of the private key, which is generated by an algorithm that selects words from a fixed vocabulary. When the user forgets the private key, the mnemonic is used to recover it.

The security of wallets focus on three essential links: the creation, storage, and use of private keys. The storage security of private keys can be strengthened through local storage. The security recovery of private keys can be enhanced through secret sharing and TTP. The secure usage of private keys can be achieved through multi-factor authentication. Non-custodial wallets, such as MetaMask~\cite{metamask}, locally store private keys, protecting them from server owners and attackers. Through academic research, the conditions for implementing secret sharing to protect private keys have evolved from relying on TTP authentication~\cite{soltani2019practical} or permissioned blockchains~\cite{ra2020key} to utilizing single-password systems~\cite{bagherzandi2011password} and trustless environments~\cite{camenisch2014memento}. The industry has also developed security-enhanced wallets based on secret sharing, such as Zengo~\cite{zengo}. Academic research has covered different types of TTP-based wallets, e.g., \cite{he2018social}'s identity-based key encryption for software wallets and \cite{lehto2021cryptovault}'s recovery for hardware wallets, and considered factors like privacy, e.g., \cite{dai2021crsa}'s recovery scheme with privacy protection using ZKPs. Argent\cite{Argent} is one of the industry examples that utilizes TTP. Multi-factor authentication, including biometric features~\cite{csahan2019multi} like fingerprint~\cite{benli2017biowallet,aydar2019private}, iris, pulse~\cite{jagadeesan2010secured}, and behavioral features like mouse behavior~\cite{hu2020securing}, can help to verify the identity of user.


\smallskip
\noindent\textbf{Oracle.} Oracle provides external data sources for smart contracts on the blockchain, supplying them with data information. 
The oracle retrieves the data from off chain data providers, typically nodes within the blockchain network, who fetch data from various public sources. The data is then sent to smart contracts of the oracle, which tasks such as packaging, verification, and cleansing of the received data. Finally, the oracle submits the updated data, allowing the user or smart contract that initiated the request to obtain.

\smallskip
\noindent\textbf{Stablecoin.} Stablecoins can be formed through various methods, including off-chain reserves or on-chain collateralization. Stablecoins circulate similarly to traditional finance systems, involving reserve, issuance, and
other essential links. 
Off-chain reserved stablecoins are backed by fiat or assets like gold. Maintaining transparency and integrity of reserve assets ensures a 1:1 collateralization ratio between stablecoins and backing assets. However, these stablecoins carry risks due to centralized reserves and third-party audits. In contrast, on-chain reserve stablecoins and algorithmic stablecoins use digital assets as collateral or eliminate collateralization altogether. They are created through a transparent on-chain process with different price stabilization mechanisms. Despite their advantages, some on-chain stablecoins are prone to downfall caused by a death spiral during crises.

\smallskip
\noindent\textbf{Lending.} Decentralized lending protocols typically involve collateralization, lending, and liquidation. 
Users provide digital assets as collateral, which are aggregated into a pool that forms a reserve used for redemption. The smart contracts issue credential tokens to users, which can be used for redemption. Users' credit for borrowing is based on the liquidity they provide, and the floating or fixed borrowing rate is determined by an interest rate contract that adjusts based on supply-borrowing dynamics according to specific interest rate models. Liquidation is triggered when a user’s debts exceed the borrowing capacity, and any participant can compete to liquidate debts and earn rewards. Some protocols distribute governance tokens to users to incentivize participation.

\begin{figure}[t]
    \centering
    \includegraphics[width=0.95\linewidth]{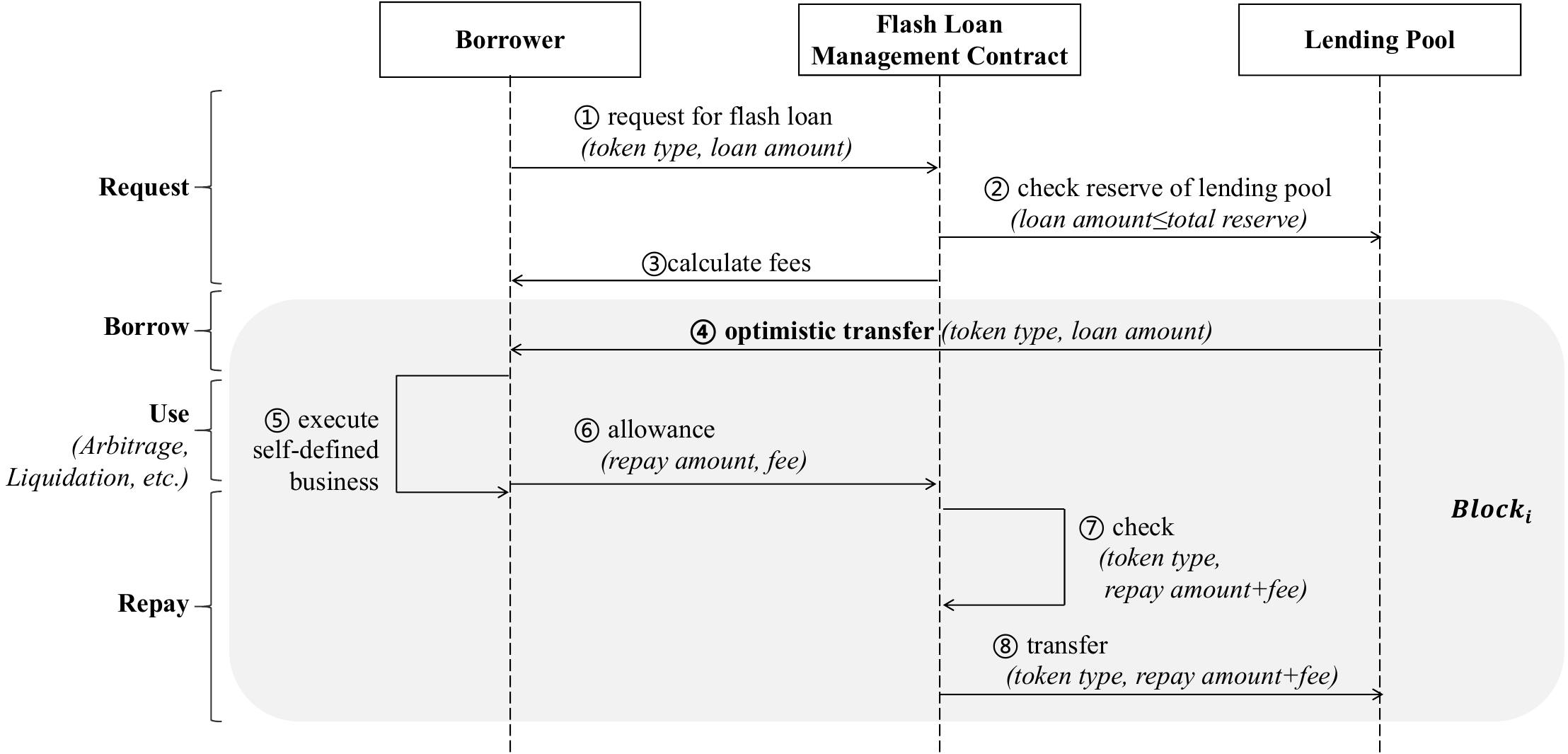}
    \caption{Flash Loan Workflow}
    \label{fig:fls}
\end{figure}

\smallskip
\noindent\textbf{Flash Loan.}
The workflow of flash loans or flash swaps is illustrated in Figure~\ref{fig:fls}.

\begin{figure}[!hbt]
    \centering
    \includegraphics[width=0.95\linewidth]{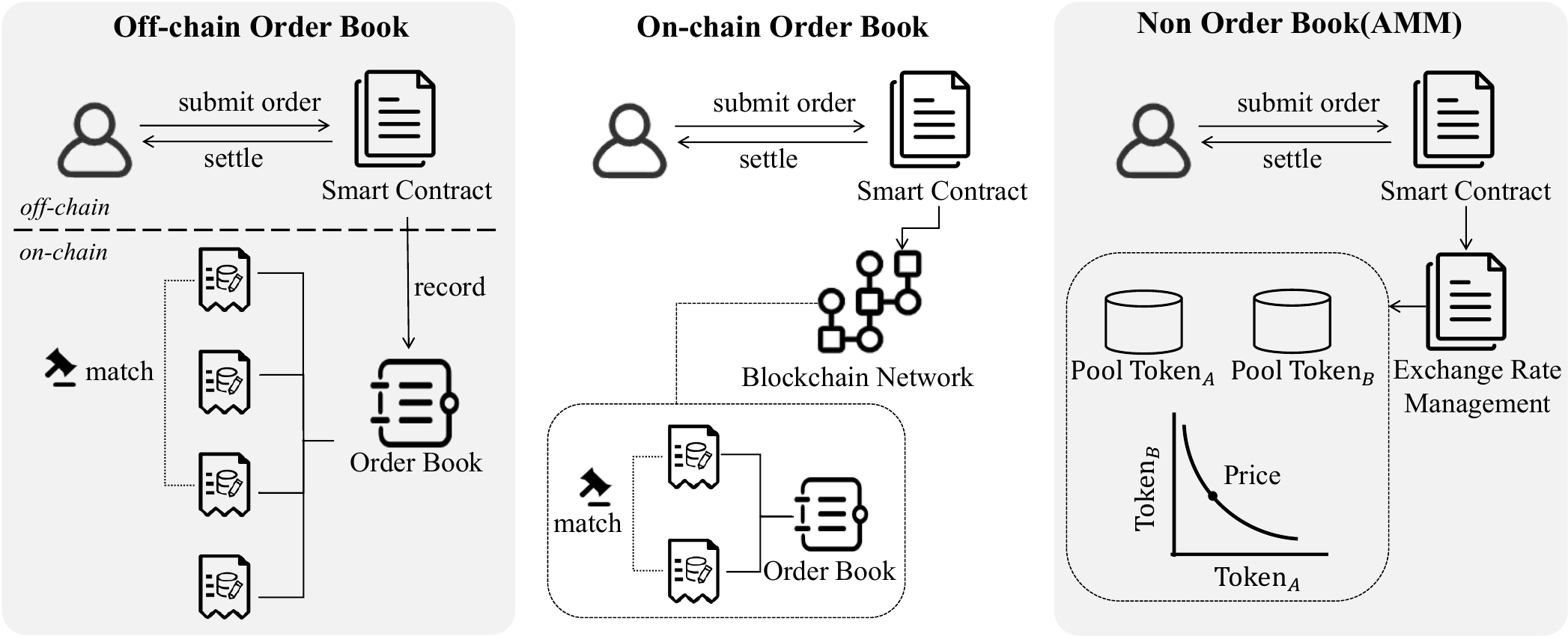}
    \caption{DEX Implementation Models}
    \label{fig:dex}
\end{figure}

\smallskip
\noindent\textbf{Exchange.} DEX can be divided into different models based on the implementation of trading pair discovery and order matching (cf. Figure~\ref{fig:dex}). %
Some DEXs use order book, where orders are recorded in an order book, and transactions are aggregated using principles of high and low bids and time order. DEXs using on-chain order books maintain order books at each node, with orders submitted to smart contracts and broadcasted to the network. When receiving the order, the node records and matches the prices and automatically executes the trade. The
discovery of transactions in this model is limited by network performance. The off-chain order book model is similar to traditional exchanges, where the exchange maintains an order book and matches them off-chain. Several DEXs innovate the non-order book model. Two methods are (i) the establishment of a reserve pool, and (ii) the use of the AMM mode, which calculates the exchange rate between two or more assets according to specific algorithms, providing the quotation between assets at any time. Both sides of AMM trades interact with on-chain liquidity pools that allow users to seamlessly switch between tokens. Liquidity providers earn income based on the percentage of their contribution to the pool. The core of AMM lies in various exchange rate algorithms, including constant mean, constant product, dynamic weighting, and constant sum.

\end{document}